\documentclass[trackchanges,twocolumn]{aastex701}

\usepackage[T1]{fontenc}
\usepackage{ae,aecompl}

\usepackage{subfigure}

\usepackage{footnote,graphicx,natbib,color,multirow,amsmath,url}

\definecolor{titlecol}{rgb}{0,0,1}
\definecolor{titlecol2}{rgb}{0,0.4,0}
\definecolor{titlecol3}{rgb}{0,0.5,0.8}
\definecolor{titlecol4}{rgb}{0.039,0.361,0.569} 
\definecolor{titlecol5}{rgb}{0.03,0.55,0.319}

\definecolor{pinknote}{rgb}{1.0, 0.0, 0.5} 
\definecolor{orangenote}{rgb}{1.0, 0.5, 0.0}

\newcommand\eg    {\emph{e.g.,}~}
\newcommand\ie    {\emph{i.e.,}~}

\newcommand\masscomp    {\textsc{mass-complete}}
\newcommand\barmass    {\textsc{bar-mass}}
\newcommand\nobarmass   {\textsc{no-bar-mass}}
\newcommand\barmasswt  {\textsc{bar-mass-wt}}
\newcommand\nobarmasswt {\textsc{no-bar-mass-wt}}

\newcommand\mstar    {$M_\ast$}
\newcommand\mmstar {M_\ast}

\newcommand\mmsun	{\rm{M}_{\odot}}
\newcommand\massz      {$(\mmstar, z)$}

\newcommand\lmstartxt  {\log \left( \mmstar / \mmsun \right)}

\newcommand\pbar         {$p_{\rm bar}$}
\newcommand\mpbar        {p_{\rm bar}}

\newcommand\pfeat        {$p_{\rm features}$}
\newcommand\mpfeat       {p_{\rm features}}

\newcommand\mpnotEO      {p_{\rm not-edge-on}}

\newcommand\lbar        {$L_{\rm bar}$}
\newcommand\mlbar        {L_{\rm bar}}
\newcommand\wbar        {$W_{\rm bar}$}
\newcommand\mwbar        {W_{\rm bar}}

\newcommand\rflux        {$r_{90}$}
\newcommand\mrflux        {r_{90}}

\newcommand\ruse         {\rflux}
\newcommand\mruse        {\mrflux}

\newcommand\rrel         {$\ell_{\rm rel}$}
\newcommand\mrrel         {\ell_{\rm rel}}
\newcommand\wrel         {$\omega_{\rm rel}$}
\newcommand\mwrel         {\omega_{\rm rel}}

\newcommand\lwrat        {$\mlbar / \mwbar$}

\newcommand\magacsi       {$I_{AB, F814W}$}
\newcommand\mmagacsi       {I_{AB, F814W}}

\newcommand\pbarthresh  {0.5}

\newcommand\mpks          {p_{\rm KS}}

\newcommand\mSFRonseq  {SFR_{\rm seq,mid}}
\newcommand\SFRonseq  {$SFR_{\rm seq,mid}$}

\newcommand\mdlogsfr  {\delta \log SFR}
\newcommand\dlogsfr   {$\delta \log SFR$}

\renewcommand\lesssim{\mathrel{\hbox{\rlap{\hbox{\lower3pt\hbox{$\sim$}}}\hbox{\raise2pt\hbox{$<$}}}}}
\renewcommand\gtrsim{\mathrel{\hbox{\rlap{\hbox{\lower3pt\hbox{$\sim$}}}\hbox{\raise2pt\hbox{$>$}}}}}

\newcommand\ngzblfull               {8,230}
\newcommand\ngzblgzhcosmosfull      {5,965}
\newcommand\nbarredtotgzblfull      {1,457}
\newcommand\nbarredtotgzhcosmos     {1,083}
\newcommand\allcosmosokgals       {175,615}
\newcommand\gzhcosmosokgals        {67,772}
\newcommand\gzblcosmosokgals        {5,245}
\newcommand\gzblcosmosokbarredgals    {940}
\newcommand\gzhcosmoscleannobardisk {5,777}
\newcommand\gzblcosmoscleanbardisk    {926}
\newcommand\mcgzhcosmosdiskall      {6,683}
\newcommand\mcgzhcosmosdisknobar    {3,565}
\newcommand\mcgzhcosmosdiskbar        {624}
\newcommand\mcgzhcosmosdisknobarwt    {641.8}
\newcommand\mcgzhcosmosdiskbarwt      {624.0}
\newcommand\mcgzhcosmosdiskbarres     {484}
\newcommand\galswbarmarks             {7,608}

\newcommand\barmassLmedres           {7.0}    

\newcommand\rrelmedzOneres             {0.407}  
\newcommand\rrelmedzOneDlores          {0.006}
\newcommand\rrelmedzOneDhires          {0.004}

\newcommand\rrelmedzFourres            {0.403}  
\newcommand\rrelmedzFourDlores         {0.017}
\newcommand\rrelmedzFourDhires         {0.010}

\newcommand\wrelmedzOneres             {0.127}  
\newcommand\wrelmedzOneDlores          {0.0005}  
\newcommand\wrelmedzOneDhires          {0.004}

\newcommand\wrelmedzFourres            {0.135}  
\newcommand\wrelmedzFourDlores         {0.004}
\newcommand\wrelmedzFourDhires         {0.002}

\newcommand\lwratmedzOneres             {3.12}  
\newcommand\lwratmedzOneDlores          {0.02}  
\newcommand\lwratmedzOneDhires          {0.04}

\newcommand\lwratmedzFourres            {2.70}  
\newcommand\lwratmedzFourDlores         {0.02}
\newcommand\lwratmedzFourDhires         {0.04}

\newcommand\LmedMOneres                 {5.91}  
\newcommand\LmedMOneDlores              {0.13}  
\newcommand\LmedMOneDhires              {0.13}

\newcommand\LmedMFourres                {8.10}  
\newcommand\LmedMFourDlores             {0.10}
\newcommand\LmedMFourDhires             {0.33}

\newcommand\lwratmedMOneres             {3.00}  
\newcommand\lwratmedMOneDlores          {0.04}  
\newcommand\lwratmedMOneDhires          {0.09}

\newcommand\lwratmedMFourres            {2.94}  
\newcommand\lwratmedMFourDlores         {0.01}
\newcommand\lwratmedMFourDhires         {0.04}

\newcommand\LmedSFROneres               {7.30}  
\newcommand\LmedSFROneDlores            {0.32}  %
\newcommand\LmedSFROneDhires            {0.21}

\newcommand\LmedSFRFourres              {6.39}  
\newcommand\LmedSFRFourDlores           {0.04}
\newcommand\LmedSFRFourDhires           {0.06}

\newcommand\rrelmedSFROneres            {0.425}  
\newcommand\rrelmedSFROneDlores         {0.001}  
\newcommand\rrelmedSFROneDhires         {0.005}

\newcommand\rrelmedSFRFourres           {0.377}  
\newcommand\rrelmedSFRFourDlores        {0.009}
\newcommand\rrelmedSFRFourDhires        {0.003}

\newcommand\wrelmedSFROneres            {0.142}  
\newcommand\wrelmedSFROneDlores         {0.001}  
\newcommand\wrelmedSFROneDhires         {0.004}

\newcommand\wrelmedSFRFourres           {0.119}  
\newcommand\wrelmedSFRFourDlores        {0.001}
\newcommand\wrelmedSFRFourDhires        {0.002}

\newcommand\lwratmedSFROneres           {2.82}  
\newcommand\lwratmedSFROneDlores        {0.09}  
\newcommand\lwratmedSFROneDhires        {0.08}

\newcommand\lwratmedSFRFourres          {3.01}  
\newcommand\lwratmedSFRFourDlores       {0.03}
\newcommand\lwratmedSFRFourDhires       {0.20}

\newcommand\lwratmedOnSeqzOneres        {3.05}
\newcommand\lwratmedOnSeqzOneDlores     {0.04}
\newcommand\lwratmedOnSeqzOneDhires     {0.04}

\newcommand\lwratmedSubSeqzOneres       {3.37}
\newcommand\lwratmedSubSeqzOneDlores    {0.02}
\newcommand\lwratmedSubSeqzOneDhires    {0.01}

\begin{document}

\title[Galaxy Zoo: Bar Lengths]{Galaxy Zoo Bar Lengths: A Catalogue of Measurements from \emph{Hubble Space Telescope} Images and the Evolution of Galactic Bar Structure at $z < 1$}

\author[orcid=0000-0002-3472-2453,sname='Hutchinson-Smith']{Tenley Hutchinson-Smith}
\affiliation{Department of Astronomy and Astrophysics, University of California, Santa Cruz, 1156 High Street, Santa Cruz, CA 96054, USA}
\email[show]{tenley@ucsc.edu}  

\author[orcid=0000-0001-5882-3323,sname='Simmons']{Brooke D. Simmons} 
\affiliation{Department of Physics, Lancaster University, Lancaster LA1 4YB, UK}
\email{b.simmons@lancaster.ac.uk}

\author[orcid=0000-0003-0846-9578,sname=Masters]{Karen L. Masters}
\affiliation{Department of Physics and Astronomy, Haverford College, 370 Lancaster Ave, Haverford, PA 19041, USA}
\email{fakeemail3@google.com}

\author[0000-0002-2583-5894,sname=Coil]{Alison Coil}
\affiliation{Department of Astronomy and Astrophysics, University of California, San Diego, 9500 Gilman Drive, La Jolla, CA 92093, USA}
\email{acoil@ucsd.edu}

\author[0000-0002-3887-6433,sname=Garland]{Izzy Garland}
\affiliation{Department of Theoretical Physics and Astrophysics, Faculty of Science, Masaryk University, Kotl\'{a}\v{r}sk\'{a} 2, Brno, 611 37, Czech Republic}
\affiliation{Department of Physics, Lancaster University, Lancaster LA1 4YB, UK}
\email{253558@muni.cz}

\author[00000-0002-6851-9613,sname=G\'eron]{Tobias G\'eron}
\affiliation{Dunlap Institute for Astronomy and Astrophysics, University of Toronto, 50 St. George Street, Toronto, ON M5S 3H4, Canada}
\email{tobias.geron@utoronto.ca}

\author[0000-0001-8010-8879,sname=Kruk]{Sandor Kruk}
\affiliation{ESA, European Space Astronomy Centre (ESAC), Camino Bajo del Castillo s/n, 28691, Villanueva de la Ca\~nada, Spain}
\email{Sandor.Kruk@esa.int}

\author[0000-0001-5578-359X,sname=Lintott]{Chris Lintott}
\affiliation{Oxford Astrophysics, Denys Wilkinson Building, Keble Road, Oxford OX1 3RH, UK}
\email{chris.lintott@physics.ox.ac.uk}

\author[0000-0001-6417-7196,sname=Smethurst]{Rebecca Smethurst}
\affiliation{Oxford Astrophysics, Denys Wilkinson Building, Keble Road, Oxford OX1 3RH, UK}
\email{rebecca.smethurst@physics.ox.ac.uk}

\author[sname=Tapia]{Amauri Tapia}
\affiliation{Department of Physics, California State University, Dominguez Hills, 1000 E Victoria St, Carson CA 90747, USA}
\email{amauri.tapia@gmail.com}

\author[0000-0002-3654-3504,sname=Willett]{Kyle Willett}
\affiliation{Minnesota Institute for Astrophysics, University of Minnesota, Minneapolis, MN 55455, USA}
\email{willettk@gmail.com}

\author[0000-0003-2556-3152,sname=Baeten]{Elisabeth Baeten}
\altaffiliation{Zooniverse Volunteer}
\affiliation{Oxford Astrophysics, Denys Wilkinson Building, Keble Road, Oxford OX1 3RH, UK}
\email{els.baeten@skynet.be}

\author[sname=Beer]{Sylvia Beer}
\altaffiliation{Zooniverse Volunteer}
\affiliation{Oxford Astrophysics, Denys Wilkinson Building, Keble Road, Oxford OX1 3RH, UK}
\email{bluemagi@live.com.au}

\author[sname=Peck]{Michael L. Peck}
\altaffiliation{Zooniverse Volunteer}
\affiliation{Oxford Astrophysics, Denys Wilkinson Building, Keble Road, Oxford OX1 3RH, UK}
\email{mlpeck54@earthlink.net}

\author[sname=Wilcox]{Julianne Wilcox}
\altaffiliation{Zooniverse Volunteer}
\affiliation{Oxford Astrophysics, Denys Wilkinson Building, Keble Road, Oxford OX1 3RH, UK}
\email{julik.wilcox@gmail.com}

\begin{abstract}

Understanding the role of galactic scale bars in disk galaxy evolution requires detailed measurements of bar properties across galaxies hosting bars at many redshifts. 
We present measurements of bar lengths and widths in a sample of 8230 disk galaxies from \emph{Hubble Space Telescope} (\emph{HST}) Legacy surveys. 
The highest-redshift barred galaxies in the sample have $z \sim 3$; most have $z \leq 1$. 
Using a mass-complete sample from the COSMOS field, we examine bar properties and evolution within $0.25 < z < 1$ in galaxies with stellar mass $\log(M_{\ast}/M_{\odot}) \geq 9.5$. 
The lowest-mass galaxies in our sample have similar star formation rate (SFR) distributions whether or not they host bars. 
For galaxies with $\log(M_{\ast}/M_{\odot}) \geq 10$, barred galaxies are more likely to be quiescent or quenched, consistent with bars mainly participating in slow quenching processes. 
The median physical bar length increases with increasing stellar mass. 
Relative bar lengths and widths (as a fraction of disk radius) peak at stellar mass $\log(M_{\ast}/M_{\odot}) \sim 10.25$, and change together with mass such that the median ratio, a proxy for bar strength, does not significantly change with stellar mass. 
Bars in our sample tend to be slightly ($\approx 13$\%) weaker at higher redshift. 
Quiescent and quenching galaxies have longer and wider bars than those in galaxies on or above the star-forming sequence, especially at lower redshift and higher masses; at the low-mass end of our sample, starburst galaxies host relatively longer and stronger bars. 
Our findings are consistent with other results from studies at both higher and lower redshift, cementing the fundamental importance of bars in disk galaxy evolution.

\end{abstract}



%
%
\section{Introduction}\label{sec:intro}
%
%

Galactic-scale bars, which are found in a significant fraction of disk galaxies out to at least $z \sim 4$ 
\citep[from over 50\% at $z \leq 0.5$ to $\sim 14$\% at $2 < z < 3$; \eg][]{sheth08,LeConte2024,Geron2025}, 
seem to be engines of galaxy evolution in several ways. In simulations, the density of the host dark matter halo significantly affects the bar rotation speed \citep{debattista98}, and bars can stabilize their host disks over long evolutionary times (for reviews of bar dynamics, see \citealt{Athanassoula2013b} and \citealt{sellwood14}). Both simulations \citep[\eg][]{kraljic12} and observations \citep[\eg][]{deSa-Freitas2025} find that most bars are long-lived ($> 1$ Gyr). Bars can exchange angular momentum with other baryonic components in the galaxy, including stars and gas both inside the bar region and in galaxy rings, contributing to the formation of spiral arms and overall disk galaxy evolution \citep{sellwood81,sparke87,pfenniger91,athanassoula03b, bournaud05}. Bars may also exchange angular momentum with the halo \citep{debattista98, athanassoula03b}, although this may not be a significant effect in the latter half of the Universe \citep{perez12}. Bars also contribute to supermassive black hole growth \citep{galloway15,Garland2024}.

Observationally, bars tend to be found in redder disk galaxies \citep{nair10b,masters11a}, and barred galaxies are more gas-poor than disk galaxies at similar stellar masses and colors \citep{masters12a}. This effect may be stronger in galaxies with stronger bars, possibly because stronger bars are more efficient at angular momentum exchange and gas consumption \citep{buta05}. For example, bars are associated with increased supermassive black hole accretion \citep{galloway15,Garland2024}. The presence of a bar may also affect the break structure of the host disk, especially in systems with smaller bulges \citep{kim14}. In local galaxy studies, bars appear different in late-type and early-type disks \citep{erwin05b, laurikainen07, aguerri09, hoyle11}, and bars seem to have more local than global influence on their host galaxies \citep{seidel15}, particularly on stars and gas in the nuclear region \citep{seidel16}. In the overall disk galaxy population, the fraction of galaxies with bars varies with redshift, stellar mass, and star formation rate \citep[\eg][and references therein]{cameron10}.

Despite the fraction of bars in disk galaxies being well-studied in the local Universe and out to relatively high redshift \citep{RC3,abraham99b,jogee04,b_elmegreen04,sheth08,cameron10,nair10b,masters12a,melvin14,simmons14,LeConte2024,Guo2025,Geron2025, HuertasCompany_bars_EC2025, EspejoSalcedo2025}, detailed studies of bar length and strength have been more limited to the local Universe \citep{erwin05b,hoyle11,kim14} until relatively recently \citep{kim21,LeConte2025}. Most findings indicate no, or little, evolution in bar lengths, which may be due at higher redshift to more complex bar origins \citep{kraljic12,Fragkoudi2025}. Here we select a sample of \ngzblfull\ disk galaxies using Galaxy Zoo morphological classifications of \emph{Hubble Space Telescope} legacy surveys \citep{willett17,simmons17}, then measure and examine their detailed bar properties.  

In Section \ref{sec:gzbl} we describe the selection of a sample of disk galaxies and measurement of bar properties. In Section \ref{sec:COSMOS} we select a subset of these galaxies from the COSMOS survey and describe the ancillary data used in subsequent analysis. We present and discuss our results in Sections \ref{sec:barredunbarred} and \ref{sec:barpropgalprop}. Throughout this paper all cross-matched catalogues use the nearest positional match within $3^{\prime\prime}$, we use the AB magnitude system, and where necessary we adopt a cosmology consistent with $\Lambda$CDM, with $H_{\rm 0}=70~{\rm
km~s^{-1}}$Mpc$^{\rm -1}$, $\Omega_{\rm m}=0.3$ and $\Omega_{\rm \Lambda}=0.7$ \citep{bennett13}.

%
%
\section{Galaxy Zoo Bar Lengths}\label{sec:gzbl}
%
%

%
%
%
%
\begin{deluxetable*}{cllr}
\tablewidth{0pt}
\tablecaption{Main workflow for Galaxy Zoo Bar Lengths.
\label{table:tree}}
\tablehead{
\colhead{Task} &
\colhead{Question} &
\colhead{Responses} &
\colhead{Next} 
}
\startdata
T0  & {\it Does this galaxy have  a bar?     } & Yes   & T1         \\
        & {                                } & No    & \textbf{end}   \\
T1  & {\it If there are spiral arms, do they start at the ends of the bar?    } & Yes                    & T2   \\
        & {  } & No                     & T2   \\
        & {                    } & No spiral arms   & T2   \\
T2  & {\it If there is a ring, does it  encircle and touch the ends  of the bar?     } & Yes           & T3   \\
        & {\it     } & No            & T3    \\
        & {\it                                } &  No ring    & T3   \\
T3  & {\it  Draw 1 line through the  length and 1 line through  the width of the bar.      } & (line markings)    & \textbf{end}  \\
\enddata
\tablecomments{All classifiers answer the initial task; if a classifier indicates that a bar is present, they are asked to further classify the spiral arms (if present), ring (if present), and then to draw the length and width of the bar on the galaxy image. At each stage, classifiers have access to ``Help'' text with further description and example images.}
\end{deluxetable*}

\begin{figure*}
\begin{center}
\includegraphics[width=\textwidth]{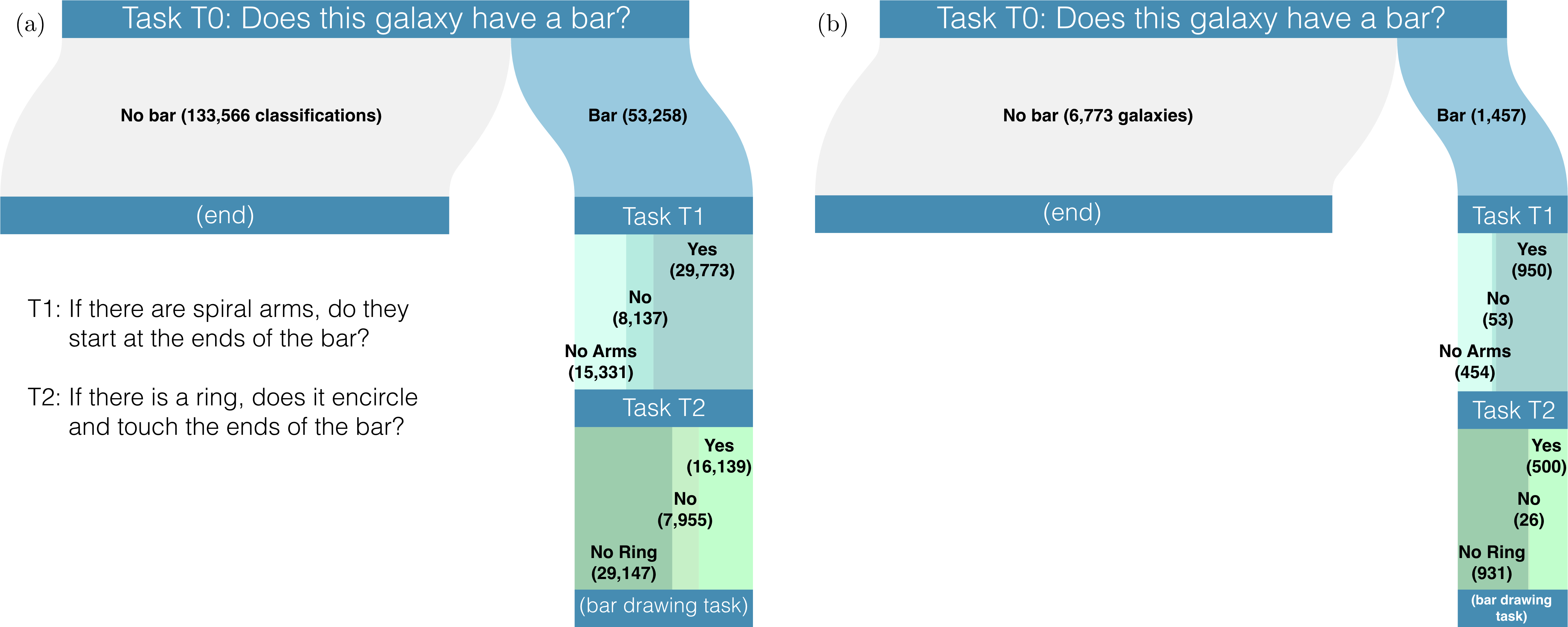}
\caption{
Demographics of the Galaxy Zoo Bar Lengths sample, based on (a) raw
classification counts and (b) plurality classification for each galaxy.
In each panel, the full classification tree (Table 1) is shown. Each
node in each diagram (dark blue horizontal bars with white text inset)
represents a task in the classification tree. The paths between tasks
represent each possible answer to the task above it, and they flow to
the next task shown to a classifier who gives that response. From the
initial task, the classification ends if the classifier indicates the
galaxy is unbarred; otherwise, they answer two further questions and
complete a drawing task, with no further branches of the classification
tree. In panel (a) on the left, the width of each path is proportional
to the number of raw classifications received for each response, for any
galaxy. In panel (b) on the right, we apply a label to each galaxy based
on the plurality (highest vote fraction) answer to each task, with ties broken by entering the
galaxy into the ``Yes'` category for that task. For example, in panel
(b), a galaxy is considered barred if at least 50\% of classifiers
answered ``Yes'' to task T0. For a task with more than 2 answer choices, the plurality need not reach the 50\% threshold. This visualization is a useful overview of
the morphologies of the disk galaxies in this dataset.
}\label{fig:sankey}
\end{center}
\end{figure*}

\begin{figure*}
\begin{center}
\includegraphics[width=0.85\textwidth]{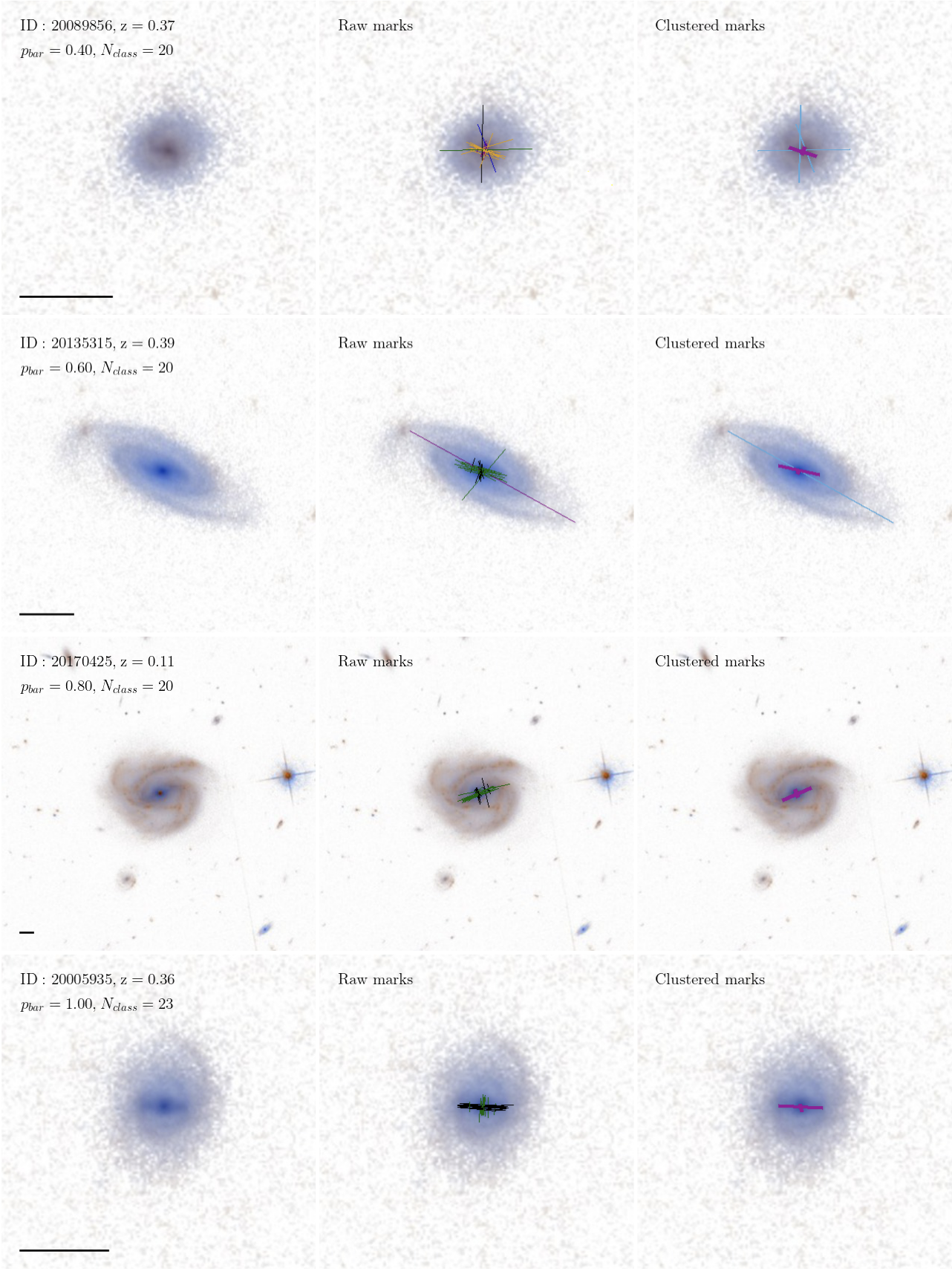}
\caption{
Example images from the Galaxy Zoo Bar Lengths project (left column),
with markings by individual classifiers shown as thin lines
(middle column) and clustered markings of bar length and width shown in
thick purple lines (right column). Marks in the middle column are color-grouped by final cluster membership. Light blue marks in the right
column indicate low-confidence clusters (typically containing outlier
marks). Agglomeration with median combination of line
parameters is robust to outliers that~\emph{do} fall within a
high-confidence cluster and often produces accurate measurements
even at values of \pbar\ below our selection threshold
(\eg top row).
}\label{fig:examplebars}
\end{center}
\end{figure*}

Galaxy Zoo Hubble \citep[GZH;][hereafter W17]{willett17} collected detailed morphological classifications of 119,849 galaxies originally imaged in 4 legacy surveys (GOODS, \citealt{giavalisco04}; GEMS, \citealt{rix04}; COSMOS, \citealt{scoville07}; AEGIS, \citealt{davis07}). The GZH project identifies galaxies with bars, but does not otherwise measure the properties of bars. The same is the case for the Galaxy Zoo CANDELS project \citep[GZC;][hereafter S17]{simmons17}, which classified galaxies at rest-frame optical wavelengths to higher redshift using images from CANDELS \citep{grogin11,koekemoer11}. We used the GZH and GZC classifications to select a sample of disk galaxies within which additional bar properties could potentially be measured. 

To select this subsample of galaxies from GZH, we used an early version of the weighted vote fractions published in W17 and S17. Following previous studies that examined barred galaxies in different Galaxy Zoo samples \citep{masters12a,melvin14,simmons14,galloway15}, we selected galaxies that classifiers had marked as likely not-edge-on disk galaxies in the first two questions in the GZH or GZC question tree (we used the same questions in each tree). Specifically, we required that a galaxy be marked as having ``features'' by at least 40\% of weighted classifiers, $\mpfeat \geq 0.4$, and that at least half of classifiers had indicated the galaxy was not edge-on, $\mpnotEO \geq 0.5$. We further selected galaxies where there was at least some indication that the galaxy might have a bar, $\mpbar \geq 0.2$.
These selections are intended to be relatively inclusive to allow for more conservative subsequent selections, while also minimising the inclusion of subjects for which asking classifiers about bar properties would be nonsensical.
These requirements resulted in the selection of \ngzblfull\ galaxies (\ngzblgzhcosmosfull\ from GZH-COSMOS, 906 from GZH-AEGIS, 610 from GZH-GEMS, 367 from GZH-GOODS, and 382 from GZC).

The detailed bar-related properties of this disk galaxy sample were collected in the Galaxy Zoo Bar Lengths (GZBL) project\footnote{\url{zooniverse.org/projects/vrooje/galaxy-zoo-bar-lengths/}}, created using the Zooniverse
Project Builder\footnote{\url{zooniverse.org/lab}}. The project presented volunteer classifiers with a galaxy image, which could be toggled to display an inverted-colormap image. These galaxy images were duplicates of the color images presented in their origin GZ projects, except further zoomed in to focus on each galaxy (by 40\% for GZH images and 60\% for GZC images). Each classifier was asked to complete up to 4 tasks for each galaxy. First, classifiers were asked to confirm whether the galaxy had a bar. If the classifier indicated the galaxy was unbarred, another galaxy was presented. If the classifier said the galaxy was barred, they were then asked further questions about the bar's relation to spiral arms (if present) and a ring (if present), and then asked to draw lines on the image to mark the bar's length and width. The classifer could optionally access ``Help'' text including examples of each galaxy type and bar markings. The full bar-length measurement workflow for the project is listed in Table \ref{table:tree}.

The GZBL project launched on the 28th of July, 2015, with the goal of collecting at least 20 classifications per galaxy image. Between the launch date and the 14th of January, 2016, GZBL collected 171,206 classifications from 2,694 registered classifiers and 730 not-logged-in IP addresses. (Each unique not-logged-in IP address was treated as a distinct classifier in our analysis.) A subsequent workflow, active between the 18th of February and the 13th of April, 2016, collected 10,952 further bar markings (task T3 only) for 1,336 subjects with marginal bar vote percentages \pbar\ for the initial task, such that further markings were required to confidently identify the consensus bar measurement. Overall, the project collected 187,431 bar-length workflow classifications of \ngzblfull\ galaxies from GZC (382 galaxies) and GZH (7,848 galaxies) over an approximately 9-month period. Of these, 607 (0.3\%) were duplicate classifications of the same  subject by the same classifier, and were removed from subsequent analysis. Excluding duplicate classifications, the average classification count per subject was 22.7; the median was 21. The overall distribution of classifications for the full sample is shown in Figure \ref{fig:sankey}.

Following the completion of the classification phase of the project for this workflow, the classifications were combined to produce an aggregated classification for each galaxy. For the question tasks (T0, T1, and T2), we computed the percentage of classifiers who gave each possible answer. For the drawing task (T3), we used agglomerative clustering to compute the bar lengths and widths. Specifically, we used a hierarchical method with the Ward distance indicator to cluster line markings based on $(x_1, y_1, r, \theta)$, where $(x_1, y_1)$ is the image coordinate of the leftmost endpoint of the line segment, $r$ is the length of the line, and $\theta$ is the angle between the line and the $+x$ axis. The distance that defines a cluster of line markings for each galaxy, $d_{\rm max}$, is defined to be the maximum distance at which the number of markings in any cluster does not exceed the number of classifiers who were presented with the marking task. 

For each cluster, we use the fraction of classifiers whose marks ended up in that cluster as a measure of cluster reliability, and consider only clusters where more than 30\% of classifiers' markings are included to be real clusters. In 81\% of galaxies where at least 6 classifiers were asked to measure bar lengths, this clustering method and reliability selection returns 2 clusters of line markings, which we take to indicate the bar length and width. In more rare cases where there are 3 or more clusters for a galaxy, we rank the clusters according to the fraction of classifiers' line marks included in the cluster, considering only the 2 most reliable clusters in further analysis.

In each cluster, we combine markings by using the median of each of the~$(x_1, y_1)$~and $(x_2, y_2)$~coordinates denoting each line endpoint. Using the median of the markings generally leads to bar lengths and widths that are within 10 degrees of perpendicular. We take the marking with the longest length to be the bar length, and the shortest to be the bar width. 

Figure \ref{fig:examplebars} shows examples of galaxies with individual classifier markings and aggregated markings. We also visually inspected all \galswbarmarks\ 
galaxies with bar markings, to assess overall data quality. The quality of marks is generally very high, consistent with a past Galaxy Zoo study of bar lengths at lower redshift using ground-based data \citep{hoyle11}. The Ward distance used in our agglomerative clustering, as well as the method chosen to determine $d_{\rm max}$, tend to include some marks which are moderate outliers in the main clusters. Taking the median of values within a cluster is far more effective at avoiding outlier influence than using mean values. 

For example, out of the 500 disk galaxies with the highest combined ``bulge prominence'' (a weighted combination of all possible responses to the Galaxy Zoo question assessing the strength of a central bulge in not-edge-on, featured galaxies; see \citealt{masters19} for its first use), approximately 25\% of images had some (typically one or two) outlier marks where classifiers missed or misinterpreted the project's explicit instructions to ignore the bulge in the bar markings. Of those, only 12 galaxies (2.4\%) had a clustered width measurement that differed from an expert (BDS) assessment of width; those differences were small ($<5$\%). Length measurements were unaffected. Given that this subset of galaxies is among the most likely to show outlier contamination, our overall outlier contamination is likely considerably below this level ($< 2$\%, based on further spot checks).

\begin{deluxetable*}{lccccccc}
\digitalasset
\tablewidth{0pt}
\tabletypesize{\scriptsize}
\tablecaption{Classifications and bar measurements for the full sample galaxies classified in the bar-length workflow in the Galaxy Zoo Bar Lengths project.
\label{table:gzblmain}}
\tablehead{
\multicolumn{1}{l}{\textsc{Data Column Name}} &
\multicolumn{6}{c}{\textsc{Subjects}} &
\multicolumn{1}{c}{}
}
\startdata
subject\_id & 466952 & 466954 & 466955 & 472198 & 472199 & 492867 & $\cdots$ \\
OBJID\_prev\_GZ & 20041145 & 20041171 & 20041181 & 50013522 & 50013555 & GDS\_18760 & $\cdots$ \\
RA & 149.9273072 & 149.9484361 & 149.9686391 & 188.9995104 & 189.1104491 & 53.0452995 & $\cdots$ \\
Dec & 1.9383032 & 1.9321485 & 1.9306812 & 62.2397809 & 62.314795 & -27.7441006 & $\cdots$ \\
scale\_arcsec\_per\_pixel & 0.01011 & 0.01011 & 0.01289 & 0.02241 & 0.01205 & 0.00849 & $\cdots$ \\
n\_class\_total                                    & 19 & 20 & 22 & 21 & 20 & 25 & $\cdots$ \\
t0\_has\_bar\_count                                & 19 & 20 & 22 & 21 & 20 & 25 & $\cdots$ \\
t0\_has\_bar\_yes\_frac                            & 0.47 & 0.55 & 0.0 & 0.24 & 0.55 & 0.12 & $\cdots$ \\
t0\_has\_bar\_no\_frac                             & 0.53 & 0.45 & 1.0 & 0.76 & 0.45 & 0.88 & $\cdots$ \\
t1\_spiral\_arms\_attached\_count                  & 9 & 11 & 0 & 5 & 11 & 3 & $\cdots$ \\
t1\_spiral\_arms\_attached\_yes\_frac              & 1.0 & 0.91 &    & 0.6 & 0.18 & 0.0 & $\cdots$ \\
t1\_spiral\_arms\_attached\_no\_frac               & 0.0 & 0.0 &    & 0.2 & 0.09 & 0.67 & $\cdots$ \\
t1\_spiral\_arms\_attached\_no\_spiral\_arms\_frac & 0.0 & 0.09 &    & 0.2 & 0.73 & 0.33 & $\cdots$ \\
t2\_ring\_attached\_count                          & 9 & 11 & 0 & 5 & 11 & 3 & $\cdots$ \\
t2\_ring\_attached\_yes\_frac                      & 0.33 & 0.36 &    & 0.6 & 0.0 & 0.33 & $\cdots$ \\
t2\_ring\_attached\_no\_frac                       & 0.11 & 0.27 &    & 0.0 & 0.09 & 0.0 & $\cdots$ \\
t2\_ring\_attached\_no\_ring\_frac                 & 0.56 & 0.36 &    & 0.4 & 0.91 & 0.67 & $\cdots$ \\
t3\_bar\_length\_class\_count                       & 10 & 10 & 0 & 5 & 11 & 3 & $\cdots$ \\
n\_lines\_highprob                                 & 3 & 2 & 0 & 2 & 2 & 2 & $\cdots$ \\
length\_p\_true\_positive                          & 0.60 & 1.0 &    & 1.0 & 0.82 & 1.0 & $\cdots$ \\
length\_x1               & 209.4 & 196.2 &    & 184.4 & 186.0 & 175.4 & $\cdots$ \\
length\_y1               & 236.2 & 262.1 &    & 209.0 & 225.0 & 188.0 & $\cdots$ \\
length\_x2               & 247.9 & 249.7 &    & 222.2 & 241.1 & 231.6 & $\cdots$ \\
length\_y2               & 157.3 & 135.8 &    & 206.0 & 232.0 & 231.0 & $\cdots$ \\
angle\_length            & -17.7 & 67.0 &    & 4.5 & -8.2 & -37.4 & $\cdots$ \\
length\_line\_dist\_pix  & 87.8 & 137.2 &    & 37.9 & 55.5 & 70.8 & $\cdots$ \\
width\_p\_true\_positive      & 0.6 & 1.0 &    & 1.0 & 0.91 & 1.0 & $\cdots$ \\
width\_x1                & 216.9 & 197.1 &    & 203.4 & 212.3 & 194.4 & $\cdots$ \\
width\_y1                & 204.5 & 186.0 &    & 201.1 & 237.5 & 216.8 & $\cdots$ \\
width\_x2                & 230.9 & 258.6 &    & 204.4 & 219.2 & 216.4 & $\cdots$ \\
width\_y2                & 186.0 & 203.5 &    & 214.1 & 218.0 & 192.0 & $\cdots$ \\
angle\_width             & 50.8 & -15.9 &    & -85.6 & 64.1 & 48.4 & $\cdots$ \\
width\_line\_dist\_pix   & 23.2 & 63.9 &    & 13.0 & 20.7 & 33.1 & $\cdots$ \\
angle\_length\_width    & 68.5  &   82.9  &     &   90.1  & 72.3  &  85.8 & $\cdots$ \\
bar\_length\_arcsec      & 0.887 & 1.387 &    & 0.850 & 0.669 & 0.601 & $\cdots$ \\
bar\_width\_arcsec       & 0.235 & 0.646 &    & 0.292 & 0.249 & 0.281 & $\cdots$ \\
i\_gal                       & 61.4  & 10.1  & 43.2 & 42.9  & 43.7  & 31.3  &  $\cdots$ \\
delta\_bar                   & 4.0   & 81.9  &      & 23.3  & -55.4 & -52.2 &  $\cdots$ \\
psi\_bar                     & 72.5  & -1.0  &      & -66.8 & 16.9  & 33.6  &  $\cdots$ \\
length\_deproject\_factor    & 1.008 & 1.016 &      & 1.066 & 1.272 & 1.110 &  $\cdots$ \\
width\_deproject\_factor     & 2.015 & 1.000 &      & 1.315 & 1.038 & 1.055 &  $\cdots$ \\
\enddata
\tablecomments{The data for each galaxy includes information that can be used to uniquely identify it in several data sets (this project as well as the earlier versions of Galaxy Zoo from which each galaxy was selected), the pixel scale of the images presented for classification in the project, the classification counts and fractions for each question in the classification tree, and information on the clustered bar measurements, including the bar length and width, where enough classifiers indicated the presence of a bar and marked its length and width to allow the agglomerative clustering to produce a result (where this is not the case, the values are blank). The complete version of this table is available in electronic form and at \url{http://data.galaxyzoo.org}. The printed table shows a transposed subset of the full table to illustrate its format and content. The complete version also includes the URLs for the galaxy images classified.}
\end{deluxetable*}

The aggregated classifications for each galaxy are shown in Table \ref{table:gzblmain}. We report the Zooniverse subject ID for each galaxy as well as its ID in the original Galaxy Zoo project from which the galaxy image was drawn (\ie either GZH or GZC). We also report the pixel scale of the classified color images and the URLs for both the standard and inverted images presented in the classification interface. The latter are included in the full version of the tables at \url{http://data.galaxyzoo.org} but are not printed in the manuscript table due to their length.

For the question tasks, the column name contains the task number and a readable description of the question and relevant quantity. For example, in the first task (T0; ``Does this galaxy have a bar?''), we report the fraction of classifiers who indicated the galaxy does have a bar (``Yes''), the fraction of classifiers who indicated the galaxy does not have a bar (``No''), and the total number of classifiers who answered the question. The three column names for this question thus are: 

\renewcommand{\labelitemi}{}
\begin{itemize}
\item
 \tt t0\_has\_bar\_yes\_frac
\item
  t0\_has\_bar\_no\_frac
\item
  t0\_has\_bar\_count
\end{itemize}

For the line measurement task (T3), we report the number of classifiers who marked the galaxy image and the number of clustered line measurements resulting from the agglomerative clustering described above. For galaxies where a bar length and width was measured, we report for both bar length and width the number of marks in that cluster as a percentage of the number of classifiers who were presented with the marking task (an indication of the likelihood the cluster is reliable), the clustered endpoint coordinates, the distance between endpoint coordinates of each line in pixels, and the angle (in degrees) between the bar length and width lines. We additionally use the pixel scale of each image to convert the bar length and width measurements to arcseconds, and report these as {\tt bar\_length\_arcsec} and {\tt bar\_width\_arcsec}. As described above, all coordinates use the median of the individual marks in the cluster.

\subsection{Correction for projection effects}

Assuming disk galaxies would be observed as circular if viewed face-on, it is possible to de-project the measured bar lengths and widths using the angle of inclination ($i$) of the disk relative to the plane of the sky, the angle of the observed disk major axis ($\theta$) and the angles of the measured bar length ($\alpha$) and width ($\beta$). Defining the relative difference between the bar length and disk major axis as $\delta \equiv \alpha - \theta$ and the relative width angle as $\psi \equiv \beta - \theta$, the factor required to correct the bar length for projection effects is
$$
f_{\rm length} = \sqrt{\cos ^2 \delta + \frac{\sin ^2 \delta}{\cos ^2 i}},
$$
and similarly for the width
$$
f_{\rm width} = \sqrt{\cos ^2 \psi + \frac{\sin ^2 \psi}{\cos ^2 i}}.
$$
In general, we use $\theta$ and $i$ values derived from position angles and axis ratios from the ACS-GC catalog \citep{griffith12}, which provides a relatively uniform set of measurements across the different fields of the \emph{HST}-ACS legacy surveys. The angles are measured from fitted light profiles \citep[using {\textsc{galfit}};][]{peng02,peng10}, which are often less affected by bright asymmetric features (such as bars) than other methods of determining angles, centroids and sizes. These values were verified by comparing them to those measured using Source Extractor \citep{bertin96} in the different surveys where available \citep{leauthaud07,caldwell08,giavalisco12}. Where the values differed substantially (3\% of galaxies with high-confidence bar markings), 
each image was visually assessed and the survey catalog values were used if the ACS-GC values were obviously incorrect. For GZC sources, which are not included in the ACS-GC, we used measurements from the survey catalogs \citep{guo13,galametz13,nayyeri17}. The values of {\tt i\_gal}, {\tt delta\_bar}, {\tt psi\_bar}, {\tt length\_deproject\_factor} and {\tt width\_deproject\_factor} are included in the GZBL catalog.

\vspace{1 cm}

\noindent We present here the full catalog of classifications and bar measurements for the GZBL project so that readers can perform their own analyses, perhaps using different thresholds than we describe below. We recommend that those pursuing their own investigations with this data require a minimum number of bar markings in order to select a sample of bars (at whatever threshold) with high confidence length and width measurements. We suggest a minimum of 6 classifiers must mark bar measurements for the clustering algorithm to work as expected. We further suggest that users set a reasonable threshold on the angle between length and width lines, \eg requiring that the lines be within $\pm 15^{\circ}$ of perpendicular.

%
\subsection{Purity and completeness}\label{sec:Purity and completeness}
%

\begin{figure}
\begin{center}
\includegraphics[width=\columnwidth]{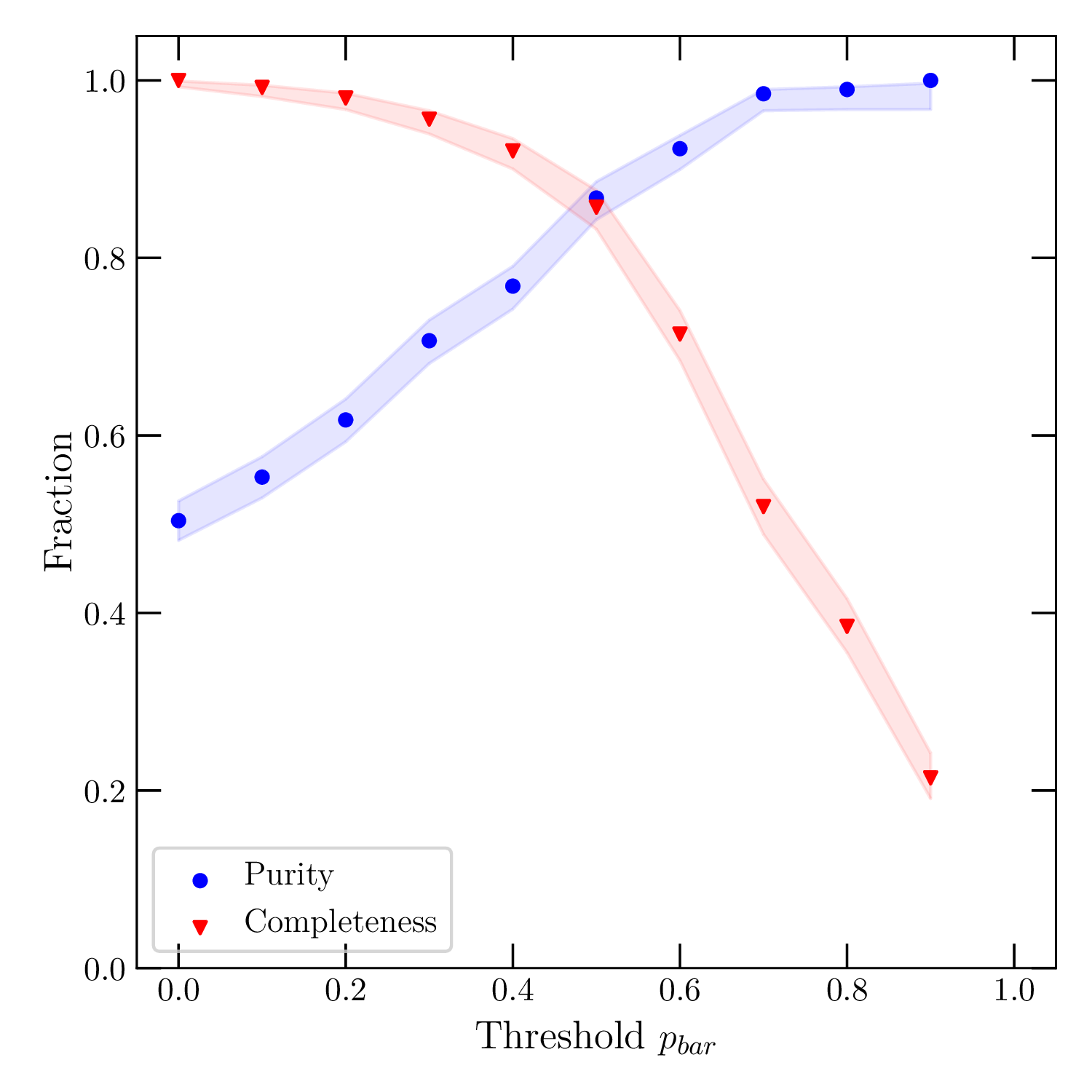}
\caption{
Purity (blue circles) and completeness (red triangles) of samples as a function of the \pbar\ threshold used to select the barred sample from within the full set of disk galaxies examined in Galaxy Zoo Bar Lengths.  As the threshold is raised, the sample becomes more pure (less contaminated by unbarred galaxies) but less complete. Considering these tradeoffs, we choose a threshold of $\mpbar = \pbarthresh$ to select a barred galaxy sample.
}
\label{fig:pc}
\end{center}
\end{figure}

\begin{figure}
\begin{center}
\includegraphics[width=\columnwidth]{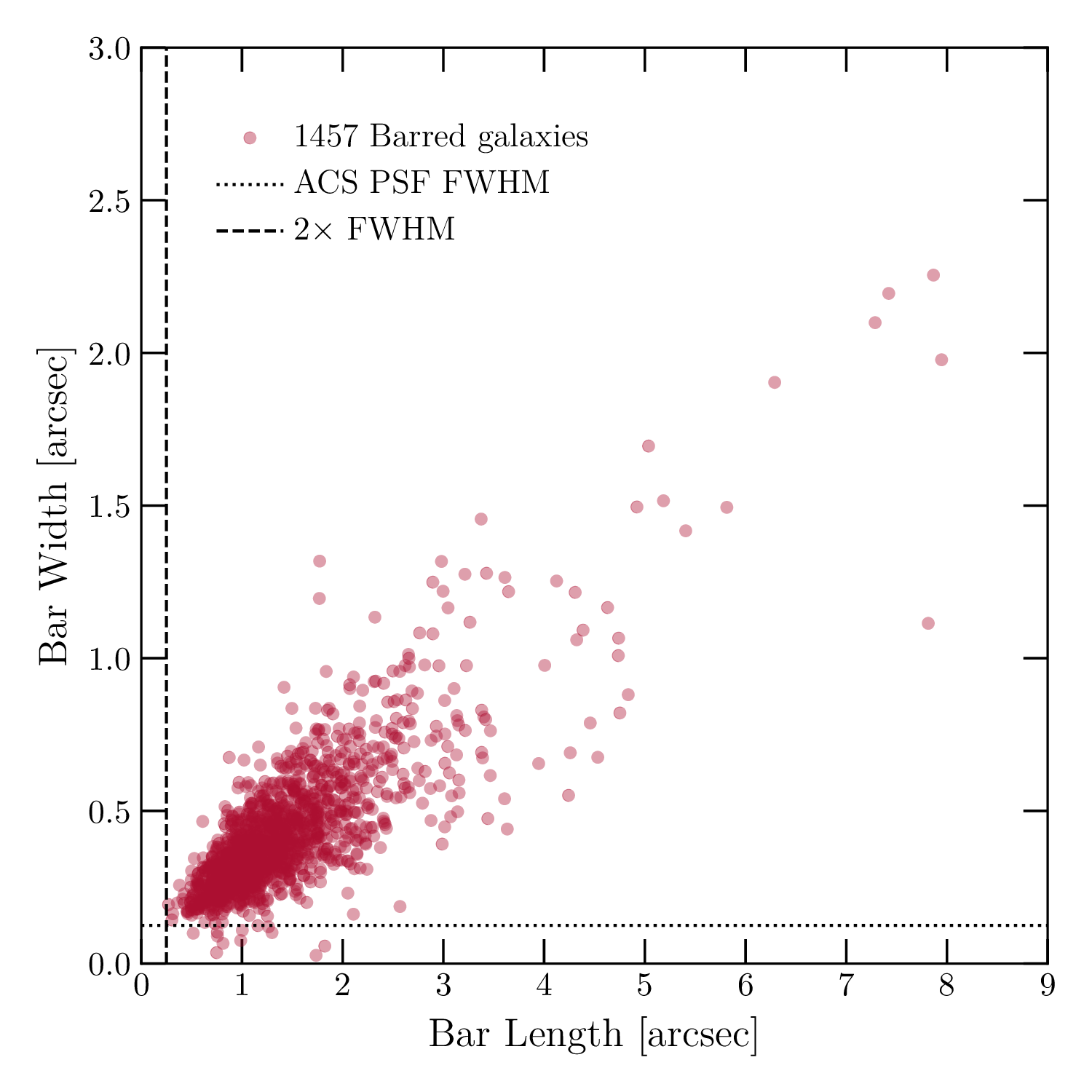}
\caption{
Bar width versus length for all bars in the Galaxy Zoo Bar Lengths project that meet the definition of barred described in Section \ref{sec:Purity and completeness}. The dotted line at constant bar width shows the FWHM size of the \emph{HST} ACS point-spread function, the minimum size we might expect to measure a bar width; the dashed line at constant length shows twice this value, the minimum size at which we would resolve a bar. Most bars are very well resolved. Bar length and width generally correlate, with a median length-width ratio of $\sim 3$. From within the full sample of barred galaxies in GZBL we further analyze the subset that have well-measured additional properties from data sets in the COSMOS field (described further in Section \ref{sec:COSMOS}).
}
\label{fig:lwall}
\end{center}
\end{figure}

In order to select barred galaxies, we choose a threshold of the bar vote percentage \pbar. In practice, this requires consideration of the trade offs between sample completeness (percentage of all barred galaxies that end up in our sample) and sample purity (percentage of galaxies in the selected sample that have bars). To assess purity and completeness in our sample, we performed expert classification on 500 galaxies selected from the full GZBL sample to uniformly span the full range of bar classifications, $0 \leq \mpbar \leq 1$. Two classifiers (THS and BDS) independently visually examined each of the 500 galaxies and assigned a classification of barred or unbarred. The independent classifications between the expert classifiers agreed very well: we report the average of their classifications, but note our analysis would not change were we to have used one or the other. 

We used these classifications to determine purity and completeness for samples selected using a threshold on \pbar. Figure \ref{fig:pc} shows both purity and completeness of barred samples as a function of the threshold value used to select the sample. We choose a threshold of $\mpbar = \pbarthresh$ when selecting a barred sample below. Using this threshold value selects a sample that is $\sim 85$\% complete and $\sim 87$\% pure. 

Applying this threshold selects \nbarredtotgzblfull\ barred galaxies. We show bar length vs bar width for all barred galaxies in GZBL in Figure \ref{fig:lwall}. The figure also shows the full width at half-maximum (FWHM) of the point-spread function (PSF) of the Advanced Camera for Surveys (ACS) in \magacsi\ as a line of constant width. The constant-length line is twice the FWHM, which indicates the minimum size at which we would expect to resolve a bar. A bar with a shorter length than this would likely not be detected. If a bar was long enough to be resolved but physically narrower than the PSF FWHM, its measured width would be limited to approximately the FWHM size. Most bars meeting the other measurement criteria described above are well resolved. 
We discuss resolution effects further in Section \ref{sec:barpropgalprop}.

The median bar length is $\sim 3$ times the bar width, regardless of how large the galaxy appears on the sky. There is substantial range of values around the median, and axis ratio is an imperfect proxy for dynamically-defined bar strength. Nevertheless, this median axis ratio is that of a strong bar \citep{abraham00,jogee04,sheth08}, consistent with previous studies using this \pbar\ threshold to select barred galaxy samples \citep{masters11a,masters12a,melvin14,galloway15}. 

Of those galaxies with $\mpbar > 0.5$ and bar lengths and widths within $\pm 15^{\circ}$ of perpendicular, the highest-redshift galaxies in the catalogue have photometric redshifts \citep{hartley13} $z \sim 2.8$, overlapping with the spectroscopically-confirmed high-$z$ bars found in surveys using \emph{JWST} \citep[\eg][]{Guo2023,LeConte2024}. Owing in large part to the wavelength coverage of the largest \emph{HST} surveys, most of the secure bar measurements in our catalogue have $z \leq 1$. Therefore, in the next section we choose a subset of the catalogue for further investigation.


\begin{deluxetable*}{lll}
\tablewidth{0pt}
\tabletypesize{\footnotesize}
\tablecaption{Selection of disks in the COSMOS subsample \label{table:discselect}}
\tablehead{
\colhead{\textsc{}} &
\colhead{\textsc{Selection}} &
\colhead{\textsc{Criteria}}
}
\startdata
1  & \tt is\_featured         & \tt gzh\_t01\_smooth\_or\_features\_a02\_features\_or\_disk\_weighted\_fraction $\geq 0.4$ \\
2  & \tt not\_clumpy          & \tt gzh\_t12\_clumpy\_a02\_no\_weighted\_fraction $\geq 0.3$ \&\& gzh\_t12\_clumpy\_total\_weight $\geq 8$ \\
3  & \tt not\_edge\_on        & \tt gzh\_t02\_edgeon\_a02\_no\_weighted\_fraction $\geq 0.5$ \&\& gzh\_t02\_edgeon\_total\_weight $\geq 8$ \\
4  & \tt not\_artifact        & \tt gzh\_t01\_smooth\_or\_features\_a03\_star\_or\_artifact\_weighted\_fraction $\leq 0.5$ \\
5  & \tt is\_odd\_only        & \tt gzh\_t06\_odd\_a01\_yes\_weighted\_fraction $\geq 0.3$ \&\& gzh\_t08\_odd\_feature\_total\_weight $\geq 6$ \&\& \\
&                             & \tt ((gzh\_t04\_spiral\_total\_weight $\geq 6$ \&\& gzh\_t03\_bar\_a01\_bar\_weighted\_fraction $< 0.3$ \&\& \\
&                             & \tt gzh\_t04\_spiral\_a01\_spiral\_weighted\_fraction $< 0.4$) || gzh\_t04\_spiral\_total\_weight $< 6$)\\
6  & \tt merger               & \tt gzh\_t08\_odd\_feature\_a06\_merger\_weighted\_fraction $\geq 0.35$\\
7  & \tt strong\_arc          & \tt gzh\_t06\_odd\_a01\_yes\_weighted\_fraction $\geq 0.5$ \&\& \\
&                             & \tt gzh\_t08\_odd\_feature\_a02\_lens\_or\_arc\_weighted\_fraction $\geq 0.3$ \\ 
8  & \tt all\_odd             & \tt is\_odd\_only \&\& (merger || strong\_arc)\\
9  & \tt clean\_disks         & \tt is\_featured \&\& not\_clumpy \&\& not\_edge\_on \&\& not\_artifact \&\& !all\_odd \\ \hline
10 & \tt disks\_unbarred      & \tt clean\_disks \&\& gzh\_t03\_bar\_a01\_bar\_weighted\_fraction $\leq 0.2$ \&\& gzh\_t03\_bar\_total\_weight $\geq 8$  \\
11 & \tt disks\_barred        & \tt clean\_disks \&\& t0\_has\_bar\_yes\_frac $\geq 0.5$ \\ \hline \hline
\enddata
\tablecomments{The complex selection criteria uses weighted vote fractions from GZH (W17) and, for selection of a barred subsample, bar vote fraction from GZBL. The final selections are made up of a number of criteria, designed to select a sample of featured, not-edge-on galaxies (1--4) and then identify different types of ``features'' which are not related to disks but rather to other disturbances to a smooth, symmetric light profile, such as mergers and gravitational lenses (5--7). Those selections are then combined to select a disk sample (8--9), from which clean samples of unbarred (10) and barred (11) galaxies are selected.}

\end{deluxetable*}

\begin{figure*}
\begin{center}
\begin{subfigure}{}
  \centering
  \includegraphics[width=.45\linewidth]{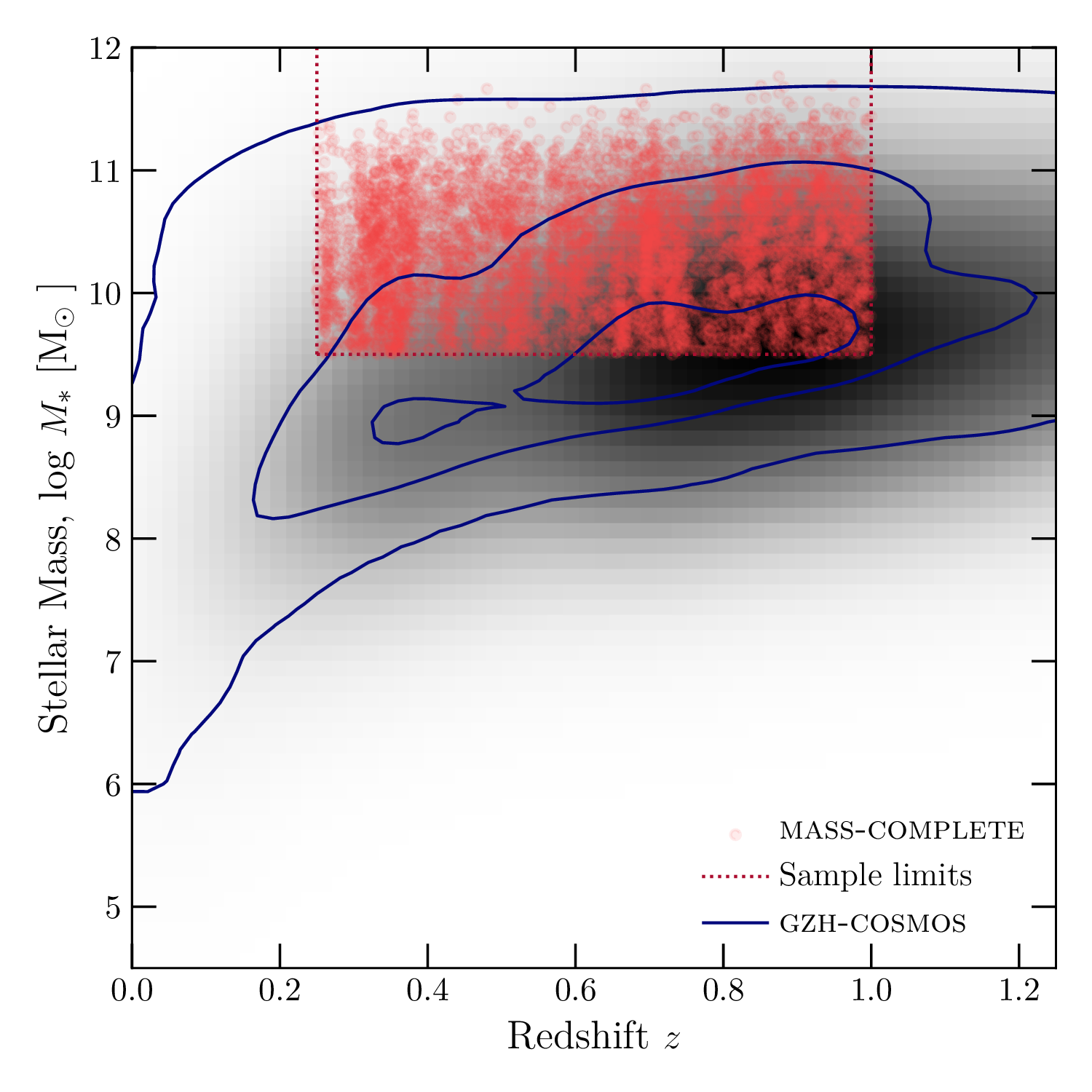}
\end{subfigure}
\begin{subfigure}{}
  \includegraphics[width=.495\linewidth]{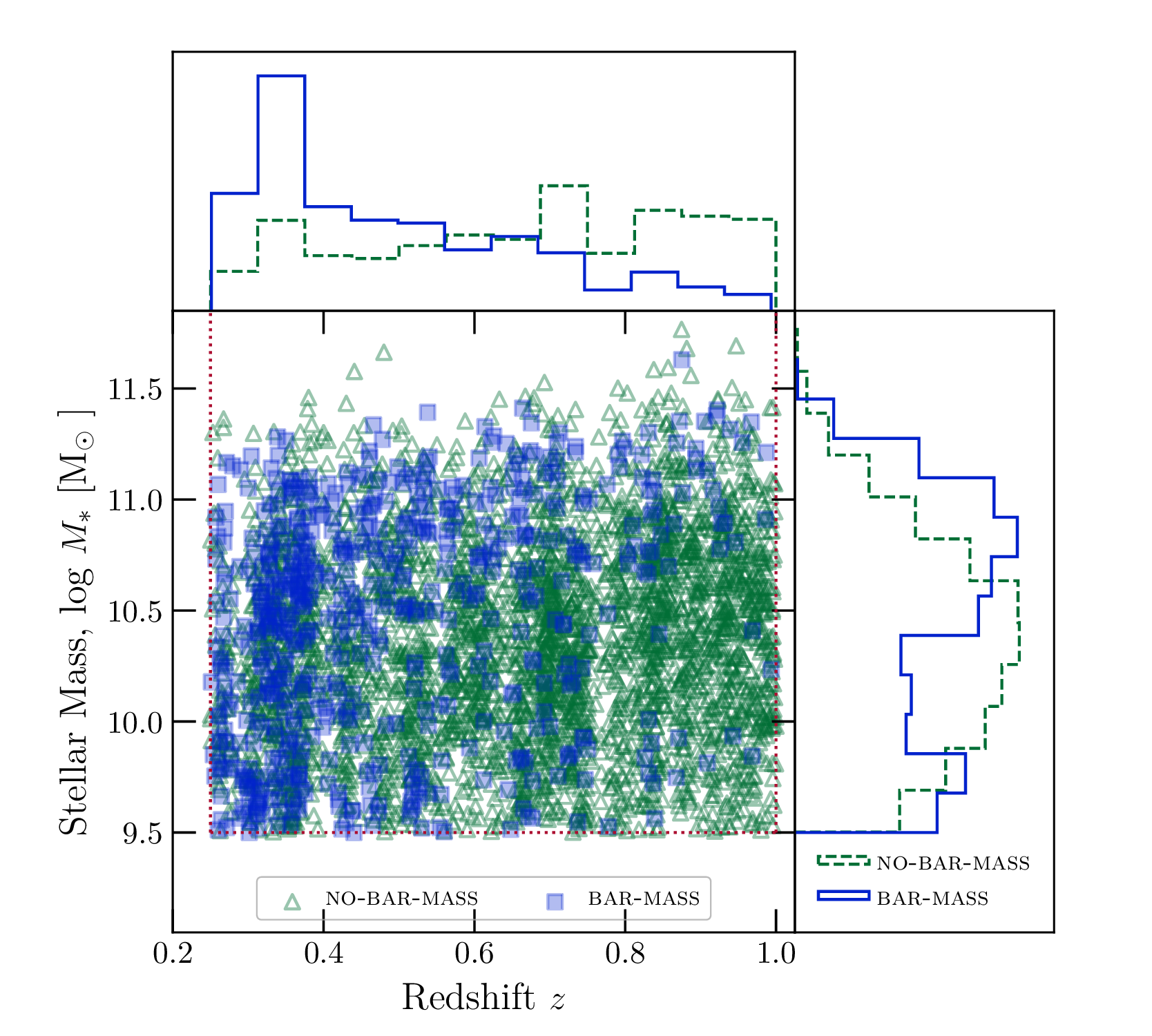}
\end{subfigure}%
\caption{
Rest-frame galaxy stellar mass versus redshift. At left, the full COSMOS sample used here \citep[Section \ref{sec:COSMOS};][]{leauthaud07,muzzin13a,muzzin13b,Weaver2022} is shown as a grayscale density plot. Solid blue contours show the subset of COSMOS galaxies examined in the GZH project (W17). From this GZH-COSMOS parent sample, disk galaxies falling within the \masscomp\ sample (limits drawn by the dotted red lines) are shown as red circles. The figure at right is zoomed to show only the limits of the \masscomp\ sample, with barred galaxies shown as filled blue squares and unbarred galaxies shown as open green triangles, and normalized mass and redshift distributions of the \barmass\ and \nobarmass\ samples shown as blue solid and green dashed lines, respectively. The \barmasswt\ and \nobarmasswt\ samples are then defined with weights such that the weighted $(M_*, z)$ distributions are statistically indistinguishable (Section \ref{sec:Mzweighting}).
}
\label{fig:zmagmass}
\end{center}
\end{figure*}

%
%
\section{COSMOS}\label{sec:COSMOS}
%
%

Above we describe and present the whole data set for GZBL, which includes \emph{HST} images from multiple surveys observing in non-uniform bands across multiple fields. For the analysis that follows we focus on the measurements made in the COSMOS ACS imaging data \citep{koekemoer07,leauthaud07}. The COSMOS ACS imaging used in the project has 0.05 arcsec/pixel resolution and covers approximately 1.8 square degrees in F814W to a depth of $\mmagacsi = 27.2$ ($5 \sigma$~point-source depth). At $z \approx 1$, the minimum resolved size of features is approximately 1 kpc. Of the \ngzblfull\ total galaxies in GZBL, \ngzblgzhcosmosfull\ are galaxies from the COSMOS ACS imaging. Of the total COSMOS subsample of disk galaxies, \nbarredtotgzhcosmos\ are barred according to our criterion.


Considerable ancillary data exists for galaxies in the COSMOS field. We make use of stellar masses (\mstar ), star formation rates (SFRs), and rest-frame colors from these catalogues. Specifically, we draw mostly from the ``classic'' COSMOS2020 catalogue \citep{Weaver2022}, using a combination of redshifts from zCOSMOS \citep{lilly07} and photometric redshifts where spectra from zCOSMOS are not available. We also use the results of spectral energy distribution (SED) fitting to source star formation rates and stellar masses, choosing the LePhare \citep{Arnouts2002,Ilbert2006} values where the quality of SED fit was good. If not, we choose the EAZY \citep{brammer08} values, if those have good fit quality. In the first instance, we draw reduced $\chi^2_{\nu}$ values from the COSMOS2020 column \texttt{lp\_chi2\_best}; in the second, we derive $\chi^2_{\nu}$ as \texttt{ez\_z\_phot\_chi2 / ez\_nusefilt}. In both cases, we find that a requirement of $\chi^2_{\nu} < 3$ adequately selects good fits. In cases where neither SED fit from COSMOS2020 meets these criteria, we use the older values from the $K_s$-selected catalogue of \citet{muzzin13a,muzzin13b}, requiring that $\rm{\tt USE} = 1$ as a criterion for data inclusion.
We further require available measurements of galaxy radii 
from the ACS-based photometric catalog \citep{leauthaud07}\footnote{That catalog was produced using a version of the COSMOS \emph{HST} imaging drizzled to 0.03~arcsec~pixel$^{-1}$ resolution. We have corrected for this difference between the two imaging pixel scales when performing analysis involving lengths originally measured in image pixels.}, and additionally apply the requirement that there be a physically reasonable range of rest-frame colors ($0.0 < U-V < 2.7$ and $-0.2 < V - J < 2.3$). Note that rest-frame $UVJ$ colors are only available from the EAZY SED fits, so these are used if the EAZY fit met the $\chi^2_{\nu} < 3$ threshold, and otherwise we use the values from \citet{muzzin13a,muzzin13b}.

While choosing a single catalogue and fit method would have been simpler, it would also have unnecessarily eliminated a small fraction of the data sources in COSMOS with measured bar lengths and widths. We note that the differences in redshifts, stellar masses, and SFRs between these 3 sources of SED fits to the same galaxies are generally small and lie within the typical uncertainties. For example, $\Delta z < 0.1$ for approximately 99\%, or $< 0.05$ for over 93\%, of disk galaxies within the volume limit we define below and with good fits between the 3 catalogues. The cascading selection criteria lead to us using SED fit parameters from LePhare approximately 96\% of the time, with COSMOS2020 EAZY fits supplementing another 1\% and the \citeauthor{muzzin13a} fits providing the remaining 3\%. For $UVJ$ colors, COSMOS2020 provides 94\% of rest-frame values, with \citeauthor{muzzin13a} providing the remainder. None of our qualitative results below would change if we were to make different catalogue selections within reasonable data quality thresholds.

These combined requirements select \allcosmosokgals\ well-measured galaxies at all redshifts, masses and colors in the COSMOS field. COSMOS2020 has a considerably fainter flux limit than that applied to select galaxies for inclusion in GZH ($\mmagacsi < 23.5$; see Section 2.1 of W17). The COSMOS sample with reliable masses, colors and radii includes  
\gzhcosmosokgals\ galaxies from the main GZH-COSMOS sample and \gzblcosmosokgals\ galaxies from the GZBL-COSMOS sample (of which \gzblcosmosokbarredgals\ are barred according to our criterion). As our GZBL-COSMOS sample is selected from GZH, the galaxies included in this analysis are drawn from the brighter, highly complete portion of the full COSMOS2020 data sets. Figure \ref{fig:zmagmass} shows the basic properties of the COSMOS sample and the subset analyzed in GZBL.

%
%
\subsection{A cleaner, more complete sample of disks}\label{sec:cleandiscs}
%
%

The selection of the GZBL sample from the GZH parent sample was relatively inclusive by design: while it was selected to include galaxies in which a bar might plausibly be measured, it did not exclude any galaxies that might have a high ``featured'' vote fraction \pfeat\ due to non-disk structures such as mergers, shells, or dust lanes in elliptical galaxies. For the analysis below, we use a stricter selection on disk galaxies by excluding features such as these. 
The disk selection uses the GZH classification question ``Is there anything odd?'' (task T06 in Figure 4 of W17) and its follow-up answers (task T08 in W17). These are designed to identify irregular, disturbed, merging, and gravitationally lensed galaxies, as well as dust lanes and rings. The variety of possible features and the need to remove non-disk interlopers from the sample \emph{without} removing disks themselves (which may be irregular and have features such as dust lanes and rings) led to a compound set of selection criteria. The specific criteria are given in Table \ref{table:discselect}; thresholds were chosen after extensive visual confirmation that using these values led to a clean yet relatively complete sample of disks. Our qualitative results below are not strongly dependent on the choice of threshold value, although some criteria are essential, such as ensuring spiral and bar features are not included in the general selection of ``odd'' galaxies. 

From within this cleaner disk sample in GZH-COSMOS, we selected a sample of unbarred disk galaxies using the bar question in that project, ``Is there a sign of a bar feature through the center of the galaxy?'' (task T03 in Figure 4 of W17). Specifically, we require that the weighted fraction of classifiers who answered ``Yes'' was no greater than 20\%, where at least 8 classifiers answered the question. The latter requirement is due to the branched nature of the GZH question tree and the position of the bar question as a third-tier question.
This selects \gzhcosmoscleannobardisk\ not-edge-on, unbarred disk galaxies from GZH-COSMOS. Applying the same disk selection to the GZBL-COSMOS barred sample described above yields \gzblcosmoscleanbardisk\ barred disks.

We wish to examine the evolution of bars to $z \sim 1$, which is the maximum redshift at which \emph{HST} imagery in the COSMOS field is dominated by rest-frame optical wavelengths. The redshift limit is also a key requirement for both examining star-formation properties in galaxies and collecting reliable bar measurements \citep[\eg][]{sheth08,speltincx08,melvin14}. We thus require a sample that is relatively complete in mass over a large redshift range. We therefore define a disk galaxy sample within $0.25 \leq z \leq 1$ and $ \lmstartxt \geq 9.5 $. This volume-limited, \masscomp\ sample, shown as enlarged red circles in both panels of Figure \ref{fig:zmagmass}, contains {\textbf \mcgzhcosmosdiskall\ disk galaxies} from GZH-COSMOS, of which {\textbf \mcgzhcosmosdisknobar\ are unbarred disks} (selected according to the criteria described above) and {\textbf \mcgzhcosmosdiskbar\ are barred disks} from GZBL-COSMOS. The remainder did not meet the criteria for secure classification into either the barred or unbarred sample.
We hereafter refer to these subsamples of unbarred and barred disk galaxies within the \masscomp\ sample as the \nobarmass\ and \barmass\ sample, respectively. We examine the \barmass\ sample in the majority of the analysis below. 


\begin{deluxetable*}{lrrllrrll}
\tablecaption{Sample counts, percentages, and confidence intervals on percentages for the volume-limited, weighted samples \barmasswt\ and \nobarmasswt\ in each star-forming category.
\label{table:sfrcats}}
\tablehead{
\multicolumn{1}{l}{} &
\multicolumn{4}{c}{\barmasswt} &
\multicolumn{4}{c}{\nobarmasswt}
\\
\multicolumn{1}{l}{\textsc{SF Class}} &
\multicolumn{1}{l}{Count} &
\multicolumn{1}{l}{Per cent} &
\multicolumn{1}{l}{.68 CI} &
\multicolumn{1}{l}{.95 CI} &
\multicolumn{1}{l}{Count} &
\multicolumn{1}{l}{Per cent} &
\multicolumn{1}{l}{.68 CI} &
\multicolumn{1}{l}{.95 CI}
\\
\multicolumn{1}{l}{} &
\multicolumn{2}{l}{ } &
\multicolumn{2}{c}{(per cent)} &
\multicolumn{2}{l}{ } &
\multicolumn{2}{c}{(per cent)}
}
\startdata
\textsc{Starburst}       & 87.0  & $13.9$\% & $_{-1.3}^{+1.5}$ & $_{-2.5}^{+2.9}$ & 122.5 & $19.1$\% & $_{-1.4}^{+1.6}$ & $_{-2.8}^{+3.2}$ \\
&&&&&&&&\\
\textsc{On Sequence} & 249.0 & $39.9$\% & $_{-1.9}^{+2.0}$ & $_{-3.8}^{+3.9}$ & 312.0 & $48.6$\% & $_{-2.0}^{+2.0}$ & $_{-3.8}^{+3.9}$ \\
&&&&&&&&\\
\textsc{Sub-Sequence}    & 200.0 & $32.1$\% & $_{-1.8}^{+1.9}$ & $_{-3.5}^{+3.8}$ & 157.4 & $24.5$\% & $_{-1.6}^{+1.8}$ & $_{-3.2}^{+3.5}$ \\
&&&&&&&&\\
\textsc{Quiescent-High}   & 63.0  & $10.1$\% & $_{-1.1}^{+1.3}$ & $_{-2.1}^{+2.6}$ &  40.2 &  $6.3$\% & $_{-0.8}^{+1.1}$ & $_{-1.6}^{+2.1}$ \\
&&&&&&&&\\
\textsc{Quiescent-Low}  & 25.0  &  $4.0$\% & $_{-0.6}^{+0.9}$ & $_{-1.3}^{+1.8}$ &   9.8 &  $1.5$\% & $_{-0.3}^{+0.6}$ & $_{-0.7}^{+1.3}$ \\\hline
\textsc{Quiescent}       & 88.0  & $14.1$\% & $_{-1.3}^{+1.5}$ & $_{-2.5}^{+3.0}$ &  50.0 &  $7.8$\% & $_{-0.9}^{+1.2}$ & $_{-1.8}^{+2.3}$ \\
~~~(combined) &&&&&&&&\\
\enddata
\tablecomments{See Section \ref{sec:rrel_dlogsfr} for the definition of each sample. As the samples are weighted to match their mass and redshift distributions, the count values need not be integers. Both 68\% and 95\% confidence intervals on population fractions are given, using binomial confidence intervals \citep{cameron11}. These distributions and counts are visualized in Figure \ref{fig:sfrmass_ratio}b, although in that figure and the rest of the analysis the 2 quiescent categories defined according to the method of \citet{aird19} are combined into a single category to improve counting statistics. The fractions for mass-and-redshift-matched barred and unbarred galaxies do not overlap within their 95\% confidence intervals for the On-Sequence, Sub-Sequence, or combined Quiescent star-forming categories.}
\end{deluxetable*}

%
%
\subsubsection{Controlling for mass and redshift differences}\label{sec:Mzweighting}
%
%

Figure \ref{fig:zmagmass} shows that the stellar mass distributions of the \barmass\ and \nobarmass\ samples are significantly different. This is a selection effect due to the incidence of strong bars being higher in galaxies with higher stellar mass \citep{sheth08,cameron10,masters12a,melvin14} at all redshift sampled here. In those analyses below which compare barred to unbarred galaxies, we correct for this effect and for differences in the samples' redshift distributions by applying a weight to each galaxy such that the weighted \massz\ distributions are statistically indistinguishable. We refer to the weighted samples below as the \barmasswt\ and \nobarmasswt\ samples.

The weighting method has the advantage of considering every galaxy in the volume-limited samples, with proportionate weights set to control for mass and redshift differences. Weighted sample counts are calculated by summing the assigned weights; the maximum weight for any galaxy is 1.0. As there are more unbarred than barred galaxies in all parts of \massz\ parameter space, this method does not downweight any barred galaxies. The effective sample sizes of the \barmasswt\ and \nobarmasswt\ samples are \mcgzhcosmosdiskbarwt\ and \mcgzhcosmosdisknobarwt , respectively. The median redshift and stellar mass of the matched samples are $z_{\rm med} = 0.49$ and $\mmstar{}_{\rm ,\ med} = 2.9 \times 10^{10} {\ \rm M}_\odot$.

We note that, in the analysis below, none of our qualitative results would have changed had we chosen subsets of the \nobarmass\ galaxies to match the \barmass\ \massz\ distributions instead of using weighted samples. We also verified that none of our qualitative results change if we consider a much more restrictive axis ratio such that only very face-on galaxies are included. As this needlessly reduces our sample size (by about half, depending on the threshold), we retain our full samples below.

\vspace{0.2in}

We note that the FWHM of the ACS point-spread function varies slightly in physical units over the redshift range in the \masscomp\ sample, from $0.3 \leq \rm{FWHM} \leq 0.6$ kpc. Morphological determinations are sensitive to resolved features larger than the FWHM, typically with sizes at least 2 times the FWHM size. In the analysis below, we are thus sensitive to bars with length $\gtrsim 1$ kpc at all redshifts considered. This means we are less sensitive to small bars compared to studies examining very nearby galaxies \citep[\eg ][]{erwin05b,erwin19}. This selection bias informs our mass limits above.


\subsection{Additional useful quantities: relative bar sizes and \dlogsfr}\label{sec:rrel_dlogsfr}

Having measured bar properties, selected galaxy samples from COSMOS2020 within the volume limit, and controlled for the differences in stellar mass and redshift between barred and unbarred populations, we must define a few further quantities before continuing.

Firstly, we combine the \emph{HST}-based measurements of galaxy size with the physical bar lengths (\lbar ) and widths (\wbar) to define the relative bar length (\rrel) and relative bar width (\wrel) as a fraction of disk size, 
\begin{equation}
    \mrrel \equiv \frac{\mlbar}{2 \times \mruse} {\rm\ \ \ \ \ and\ \ \ \ \ }\mwrel \equiv \frac{\mwbar}{2 \times \mruse}
\end{equation}
where \ruse\ is the 90\% $F814W$ flux radius \citep{leauthaud07} of the galaxy. This flux radius encompasses most of the disk and is far less influenced by fluctuations in central concentration (\eg due to bulges or bars) than radii which only consider smaller fractions of the galaxy light \citep{Graham2019}. Choosing \ruse\ over the more commonly-used half-light radius thus considerably reduces the possibility of introducing a correlation between relative bar size and other galaxy properties which also correlate with bulge strength (such as star-forming properties).

Secondly, we seek a parameter which quantifies a galaxy's star-forming status compared to the population. The star formation rate of a ``typical'' star-forming galaxy depends on stellar mass and redshift \citep[best seen in studies of the star-forming sequence, \eg][]{Noeske2007,Peng_Y_2010,aird19}. Over the range in \massz\ parameter space of our samples, these dependencies may complicate our analysis and wash out real trends and differences. \citet{aird19} measured the dependence of the $SFR-\mmstar$ relation on both stellar mass and redshift (their Equation 8). This allows us to predict, for each galaxy in our sample, the star formation rate that the galaxy would have, \SFRonseq, if it were in the middle of the star-forming sequence. The ratio between this hypothetical ``mid-sequence'' value and the galaxy's observed $SFR$ (or the difference, in log),
\begin{equation}
    \mdlogsfr \equiv \log SFR - \log \mSFRonseq
\end{equation}
gives a value that measures the offset between the galaxy's SFR and the star-forming sequence, and which accounts for the dependency of the sequence on both \mstar\ and $z$. Positive values of \dlogsfr\ are above the midpoint of the sequence, and negative values are below it. 

Using \dlogsfr , we also adopt the definitions of star-forming categories \{Starburst, On Sequence, Sub-Sequence, Quiescent-high, Quiescent-low\} recommended by \citet{aird19}, and apply these to both the \barmasswt\ and \nobarmasswt\ samples. Table \ref{table:sfrcats} presents weighted sample counts and fractions for both samples in all categories. In general, there are relatively few quiescent galaxies in our volume-limited, \massz-controlled samples, so we combine the two Quiescent categories into a single category for the rest of the analysis. 

\begin{figure}
\begin{center}
\includegraphics[width=0.475\textwidth]{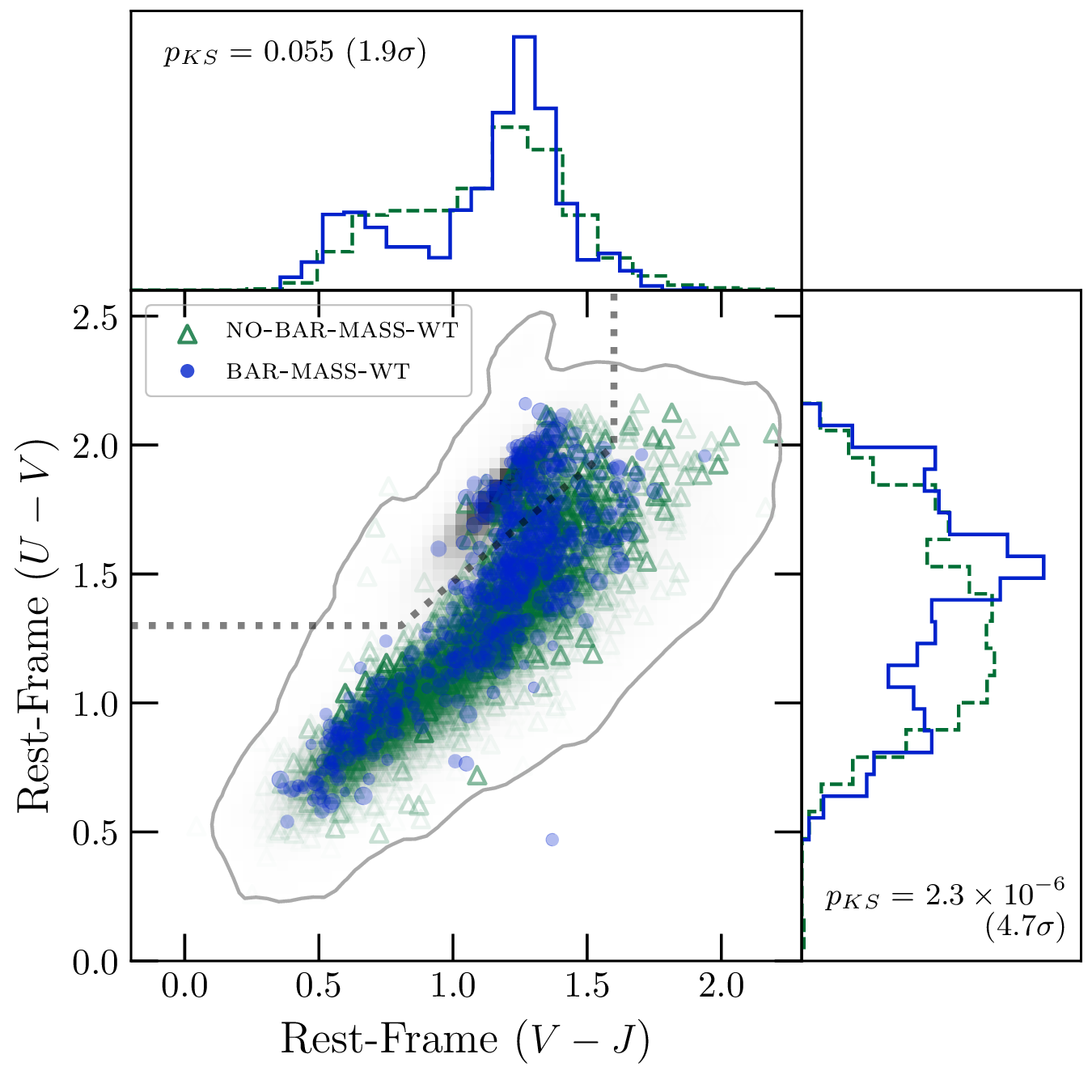}
\caption{
Rest-frame $U-V$ versus $V-J$. The gray density plot shows all galaxies (of all morphologies) COSMOS within the boundaries of the \masscomp\ sample; the gray contour encloses 99\% of COSMOS galaxies. Points show COSMOS disk galaxies analyzed by the Galaxy Zoo Bar Lengths project, split into \barmass\ (blue circles) and \nobarmass\ (green triangles) sub-samples. For both sub-samples, the opacity of each point is scaled according to its weight, such that the weighted \barmasswt\ and \nobarmasswt\ samples have statistically indistinguishable stellar mass and redshift distributions. For barred galaxies, the size of the point is proportional to the fractional bar length \rrel . The dashed line separates passively-evolving (top-left region) and star-forming galaxies \citep[from][$z \geq 0.5$ coefficients]{williams09}. Histograms of $U-V$ and $V-J$ for the \barmasswt\ (blue solid lines) and \nobarmasswt\ (green dashed lines) distributions are also shown: even controlling for stellar mass and redshift, barred and unbarred galaxies have different color distributions.
}
\label{fig:colorcolor}
\end{center}
\end{figure}

\begin{figure*}
\begin{center}
\begin{subfigure}{}
  \centering
  \includegraphics[width=0.45\textwidth]{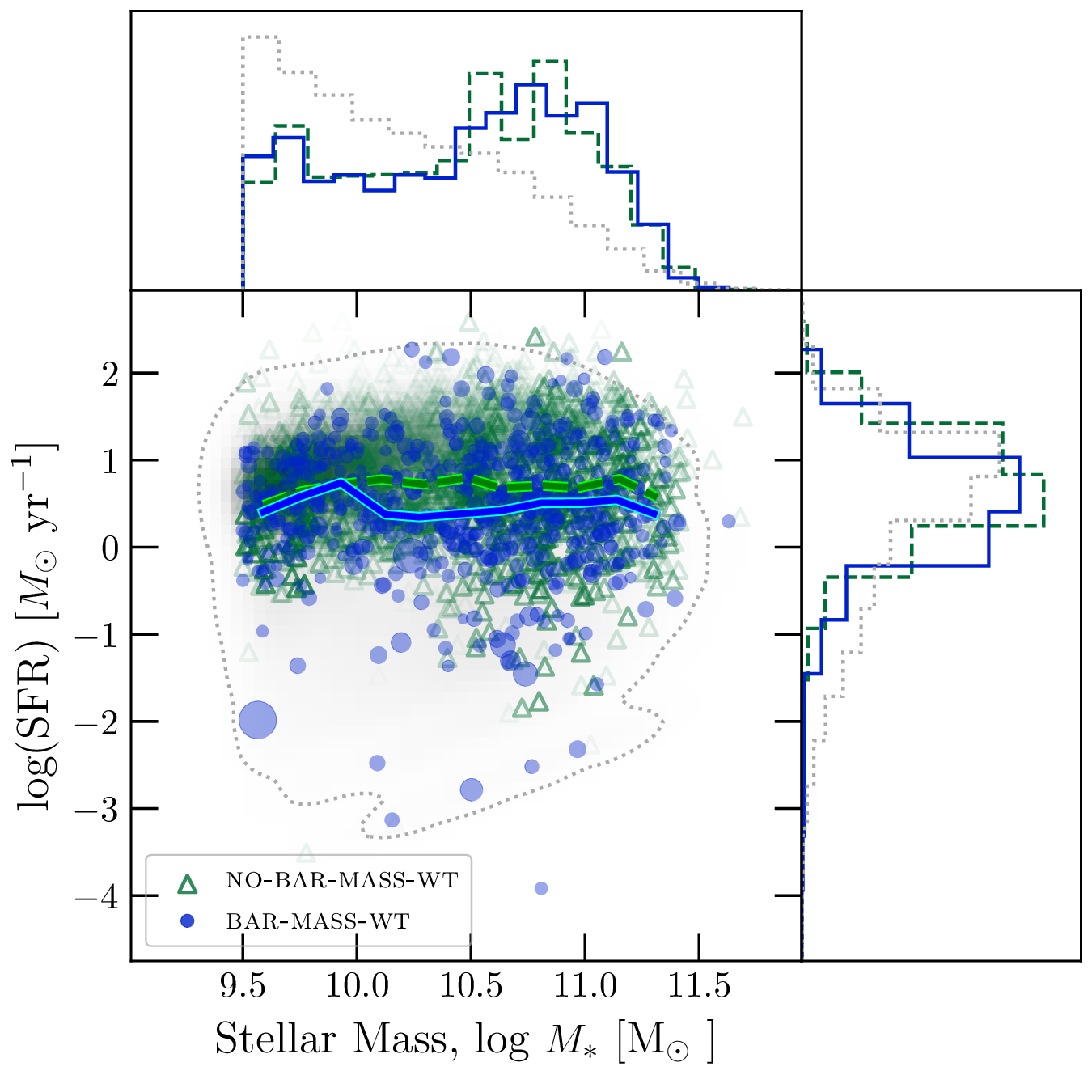}
\end{subfigure}
\begin{subfigure}{}
  \centering
  \includegraphics[width=0.45\textwidth]{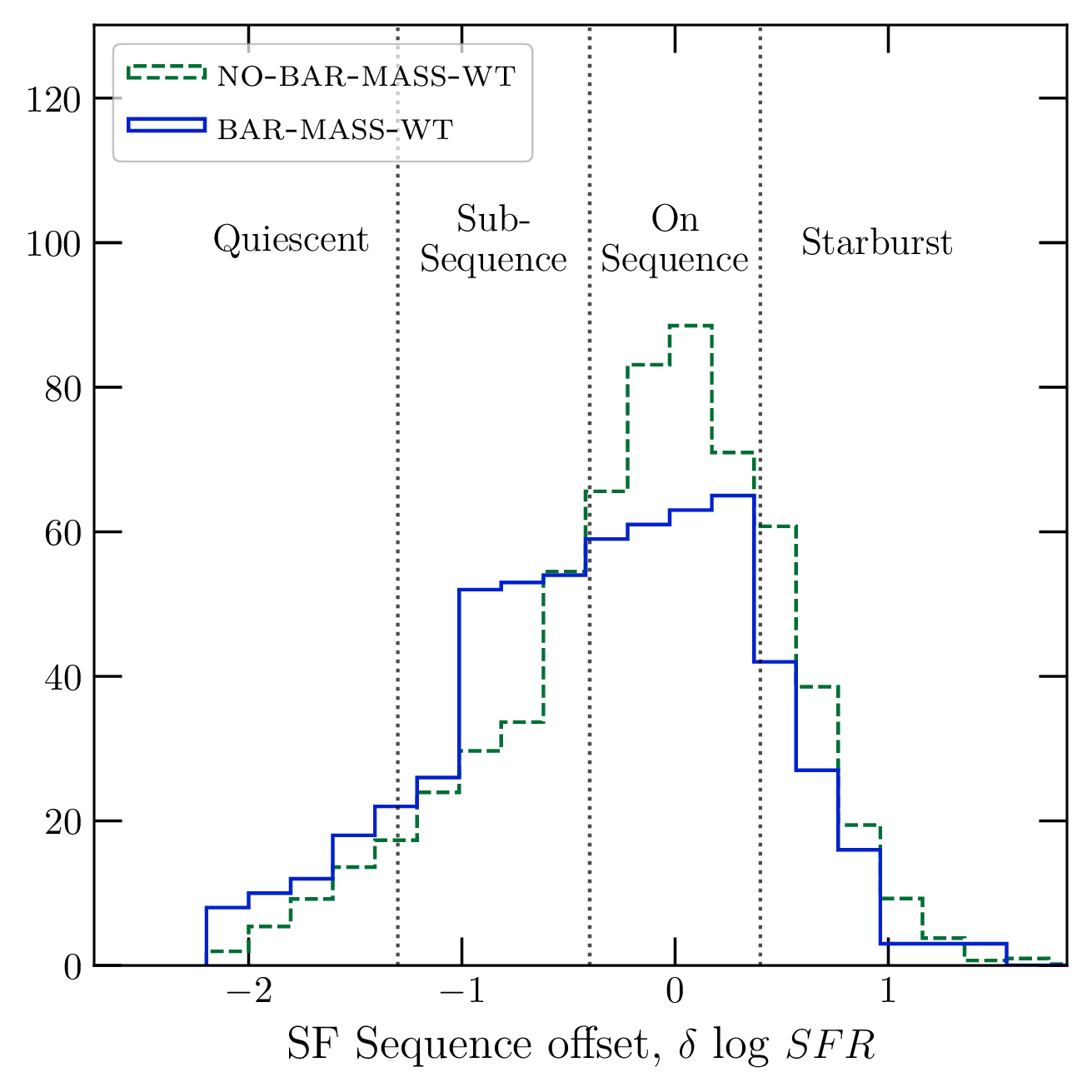}
\end{subfigure}
\caption{
Left: star formation rate versus stellar mass for barred (blue circles) and unbarred (green triangles) disk galaxies, as well as for the overall COSMOS sample within the mass and redshift limits (gray 2D histogram and outer 99\% dotted contour). The size of each barred-galaxy point is scaled to the relative bar length, as in Figure \ref{fig:colorcolor}. Each point's opacity is scaled according to its weight, such that the points represent the weighted \barmasswt\ and \nobarmasswt\ samples. Binned medians for the \barmasswt\ and \nobarmasswt\ samples are shown as blue solid and green dashed lines, respectively.  Axis histograms show weighted sample counts for barred (blue solid) and unbarred (green dashed) disks, and all galaxies (grey dotted) in the volume limit. Both samples occupy the full range of all galaxies, although there are relatively few disk galaxies with features such as spiral arms and bars that are both low-mass and off the star-forming sequence. Right: distribution of positions of \barmasswt\ and \nobarmasswt\ galaxies relative to where they would be expected if they were in the middle of the star-forming sequence at their stellar mass and redshift, according to the method of \citet{aird19}. The lowest-value bin includes galaxies with \dlogsfr\ lower than that value (highly quiescent galaxies). Generally, there is an excess of barred versus unbarred disk galaxies below the star-forming sequence. These categories will be used for analysis in sections below.}
\label{fig:sfrmass_ratio}
\end{center}
\end{figure*}

\section{Star formation Properties of barred versus unbarred galaxies}\label{sec:barredunbarred}


To analyze the distribution of star formation activity between barred and unbarred galaxies,
we use both rest-frame colors and SFR measurements, as each captures slightly different physical processes. Specifically, rest-frame colors generally reflect a combination of a galaxy's current star formation rate and longer-term star formation history, while SFRs more directly trace recent star-forming activity on shorter timescales.  

Figure \ref{fig:colorcolor} shows rest-frame $U-V$ versus $V-J$ colors for barred and unbarred galaxies in the \masscomp\ sample. 
We show the weighted samples, \barmasswt\ and \nobarmasswt\, visualising the weight of each source with the opacity of the point representing it. We adopt the definition of passively evolving versus actively star forming from \cite{williams09}, shown as dotted lines in Figure \ref{fig:colorcolor}, and additionally show the density plot and outer 99\% contour for the full COSMOS2020 data set (\ie galaxies of any morphology) within the mass and redshift limits of the \masscomp\ sample. The passive-active definition in \citeauthor{williams09} evolves slightly between $z < 0.5$ and $z \geq 0.5$. Figure \ref{fig:colorcolor} shows the diagonal separation for $z \geq 0.5$, and below we apply labels based on the definition appropriate to each galaxy's redshift.

The majority of disk galaxies in our sample are actively star forming, with a smaller fraction of disks passively evolving. Barred galaxies exhibit different rest-frame color distributions compared to unbarred counterparts, even after controlling for mass and redshift. Specifically, barred galaxies are underrepresented in intermediate $U-V$ and $V-J$ space. The differences between distributions are larger and more statistically significant in $U-V$ than $V-J$: according to a 2-sided Kolmogov-Smirnov test 
\citep[KS;][]{Kolmogorov1933,Smirnov1939,Smirnov1948,Massey1951}\footnote{Applying the KS test to weighted samples is straightforward, by measuring distribution distances using weights rather than integer counts to compute cumulative fractions. Estimating $p$ values uses weighted sample sizes, rounded to the nearest integer value. As no source in either sample has a weight above 1.0, this estimate of significance tends to be conservative.} between \barmasswt\ and \nobarmasswt, $p_{KS,\,U-V} = 2.3 \times 10^{-6}$ ($4.7 \sigma$) versus $p_{KS,\,V-J} = 0.055$ ($1.9 \sigma$). 

Using the active-versus-passive galaxy criteria from \citet{williams09} to separate actively star-forming versus passively-evolving galaxies reveals that the differences in color distributions originate in the active locus. For the passive sub-samples, we cannot rule out the null hypothesis that the barred and unbarred galaxies are drawn from the same parent sample ($p_{KS,\,U-V} = 0.98$ versus $p_{KS,\,V-J} = 0.42$ when considering only the passive subsamples at all redshifts). For the ``active'' galaxies, the color differences in $U-V$ remain statistically significant ($p_{KS,\,U-V} = 7.7 \times 10^{-4}$, $3.4 \sigma$, versus $p_{KS,\,V-J} = 0.076$, $1.8 \sigma$) despite the drop in sample size.

A star-forming galaxy's position in the active region of the $UVJ$ diagram depends on many physical properties, including stellar mass (with more massive galaxies moving towards the redder end of the region because they emit more light from previous generations of star formation),  
current SFR \citep{patel2011} and dust extinction along the line of sight \citep{williams09}. The latter would make a larger difference in the subset of galaxies which is less face-on in our samples. To assess whether orientation might drive the observed color differences, we examined a version of Figure \ref{fig:colorcolor} that included only face-on disks (assuming intrinsically circular disks and requiring axis ratios $b/a \geq 0.8$), as well as a version where axis ratio was not thresholded but instead controlled for in a 3-way ($\mmstar-z-b/a$) weighting. The same trends remained apparent; the statistical significance increased slightly (\eg from 4.7 to $4.9 \sigma$ for $U-V$ and 1.9 to $2.5 \sigma$ for $V-J$ when considering the ($\mmstar-z-b/a$)-weighted subsample), suggesting that dust-related orientation effects may play a small role in observed differences, but do not dominate the results.

The \barmasswt\ and \nobarmasswt\ samples are weighted so as to control for stellar mass and the effect of observing at different evolutionary epochs. The color difference therefore cannot be due to differences in mass or redshift between barred and unbarred samples. The declining bar fraction with increasing redshift in the overall galaxy population \citep[\eg][both of which examined bar fractions in the COSMOS-\emph{HST} field]{sheth08,melvin14} means there are fewer galaxies in either sample in the higher-redshift panel of Figure \ref{fig:colorcolor}. While this makes the $UVJ$ color gap in the active locus more visually evident in the $0.5 < z \leq 1.0$ scatter plot, the smaller sample size also decreases the statistical significance in that bin. Further examination of possible trends with redshift is warranted, and will be addressed in Section \ref{sec:barpropgalprop}. For now, we will compare the SED-fitting-based SFR properties of barred and unbarred galaxies.




The left panel of Figure \ref{fig:sfrmass_ratio} shows that both barred and unbarred disk galaxies span the full range of stellar masses and SFRs within the volume-limited COSMOS2020 sample. As discussed in Section \ref{sec:Mzweighting} above, the peak value of the mass-matched distributions at $\lmstartxt \sim 10.7$ is at least partially due to selection effects. As expected, disk galaxies undersample the quiescent portions of the parameter space compared to all galaxies in the volume, although both barred and unbarred disks are present below the star-formation sequence. Barred galaxies slightly overpopulate the portion of the diagram just below the star-forming sequence compared to their unbarred counterparts.
The binned medians in this panel show that, at stellar masses $\mmstar \gtrsim 10^{10}\ \mmsun$, the barred disk population's median SFR decreases to just below that of the unbarred disk population.


To investigate this further, the right panel of Figure \ref{fig:sfrmass_ratio} plots each galaxy's offset from the star-forming sequence (\dlogsfr), which removes the redshift and mass dependence of the primary SFR$-\mmstar$ relation. We confirm that barred galaxies are more likely to lie below the star forming sequence than unbarred galaxies. The overall distributions are significantly different ($\mpks = 1.6 \times 10^{-6}$, or $4.8 \sigma$ significance). Table \ref{table:sfrcats} shows this numerically: while the weighted sample counts in the Starburst category are consistent between barred and unbarred disks within their 95\% confidence intervals, this is not the case for any of the On Sequence, Sub-Sequence, or combined Quiescent categories. There are proportionally fewer barred galaxies compared to unbarred on the star-forming sequence, and proportionally more barred than unbarred below the sequence. 

As indicated by the left panel of Figure \ref{fig:sfrmass_ratio}, this difference is mass-dependent: the SFR distributions of barred and unbarred disk galaxies are statistically indistinguishable for the subset with $\lmstartxt \leq 10.0$ ($\mpks = 0.69$, $0.4 \sigma$), whereas for both $10.0 < \lmstartxt \leq 10.5$ and $\lmstartxt > 10.5$ the distributions are inconsistent with being drawn from the same parent distribution ($\mpks = 1.3 \times 10^{-5}$ and $2.6 \times 10^{-5}$, respectively; both are $> 4 \sigma$ significance).

This relative over-representation of barred galaxies on the low-SFR side is consistent with previous studies finding that bars play a role in accelerating the quenching process, in both simulations and observations. For example, bars are efficient engines of angular momentum transfer, driving gas inflows in simulations \citep[\eg][]{athanassoula92a, athanassoula13} which lead to central star formation and eventual quenching \citep[][]{Carles2016}.
Observationally, strong bars are associated with an overall HI gas deficiency \citep{masters12a,Fraser-McKelvie2020b}, and show ionisation properties consistent with low star-formation efficiency \citep{Krishnarao2020}. There is often a change in SFR along the bar itself \citep{Fraser-McKelvie2020a,geron24}, a finding which depends on mass in a way that is consistent with simulations \citep[][]{Carles2016} and with our results. The impact of the bar on the host galaxy increases when a strong bar is also a slow rotator \citep{geron24}. 

From the wider view offered by this sample (Figure \ref{fig:sfrmass_ratio}), bars are clearly associated with quenching of the overall galaxy, at larger stellar masses ($\mmstar \geq 10^{10}\ \mmsun$). This interpretation is supported by prior work that discusses specific pathways through the green valley for galaxies with different morphologies and structures \citep[\eg][]{Noeske2007,schawinski14,smethurst15,Bremer2018,Geron2021,Mukundan2025}. In particular, our results are consistent with past studies finding that the quenching of galaxies associated with bars is relatively slow \citep[\eg][]{Nogueira-Cavalcante2018,Kelvin2018}, causing barred disks to ``pile up'' in the green valley compared to unbarred galaxies. This, and the associated higher fraction of quenched barred galaxies, shows up in the $UVJ$ diagram (Figure \ref{fig:colorcolor}) as a decrement in the $U-V$ and $V-J$ colors just blueward of the transition to the passive region, with a corresponding increase in barred galaxies at or above (redward of) the transition, compared to the distribution of colors of unbarred disk galaxies.


\begin{figure*}
\begin{center}
\includegraphics[width=0.85\textwidth]{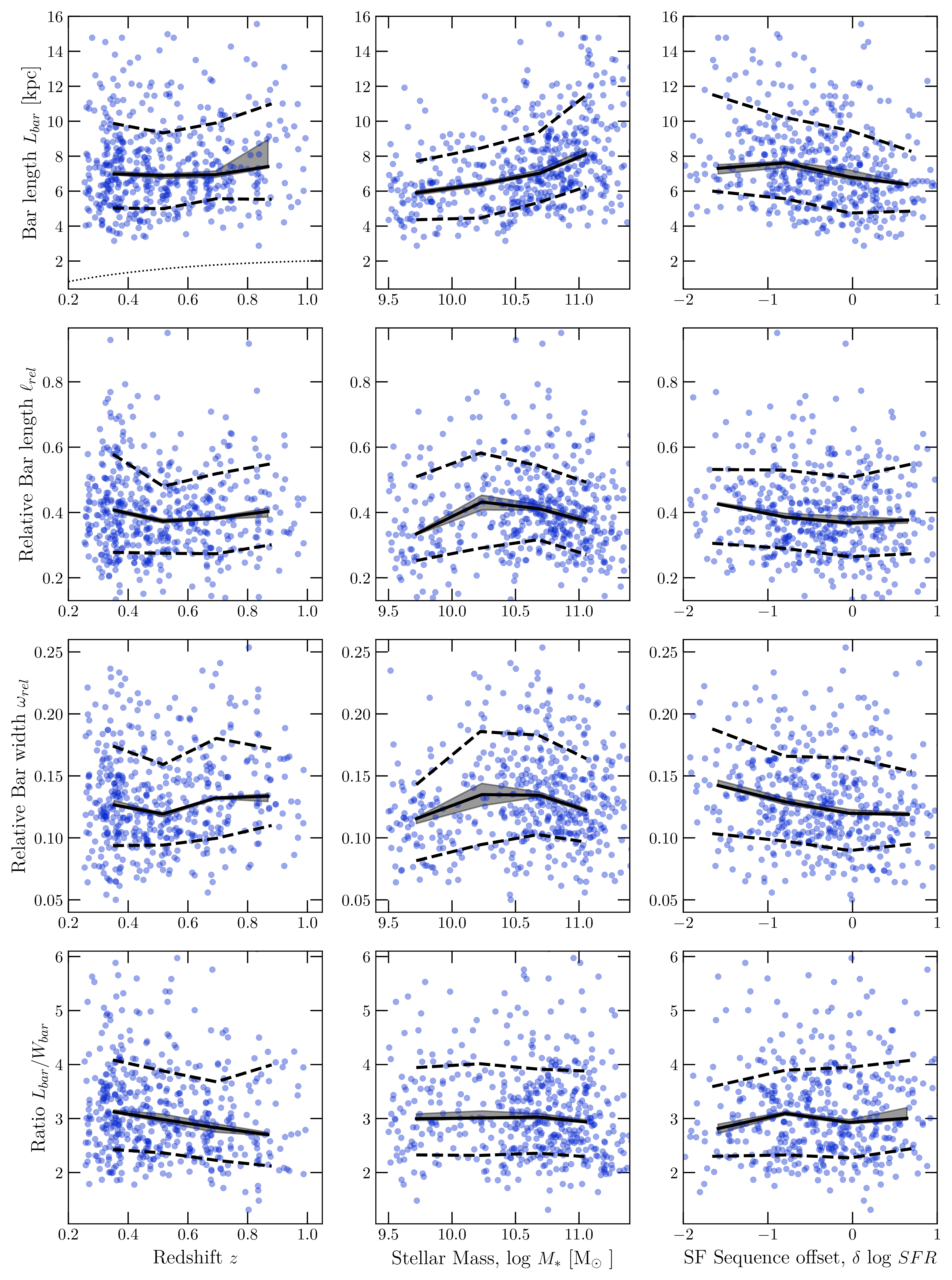}
\caption{Properties of bars in the subset of the \barmass\ sample with bar width $\mwbar > 1.5$ kpc (blue circles) versus galaxy properties. From top to bottom, $y$-axes show bar length in kpc, relative bar length \rrel , relative bar width \wrel , and length-to-width ratio \lwrat\ (a measure of bar strength). From left to right, $x$-axes show redshift $z$, stellar mass \mstar, and offset from the star-forming sequence \dlogsfr. Solid lines show binned median bar property values, with 95\% confidence intervals on the median, estimated by varying binning and resampling 1,000 times within $\pm 0.1$ in each $x$-axis value, shown as gray shaded regions. Dashed lines enclose the central 68\% of galaxies in each bin. In the top left panel, the dotted line shows the evolution of $2 \times$\,PSF FWHM (\ie the minimum resolvable bar length) with redshift. There are some statistically significant trends in the median values, although these differences are much smaller than the width of the overall population distributions. }
\label{fig:overallbarprops} 
\end{center}
\end{figure*}

\section{Overall properties of bars in galaxies at $0.25 < z < 1$ }\label{sec:barpropgalprop}

The goal of this section is to examine the structural properties of bars -- including bar length \lbar, relative bar length \rrel, relative bar width \wrel, and length-to-width ratio \lwrat \ (an imperfect measure of bar strength) -- and their connection to host galaxy properties such as redshift, stellar mass, and SFR. 

In Section \ref{sec:Purity and completeness}, we noted that a small fraction of clearly detected bars have width measurements indicating the physical widths may not be well resolved (also see Figure \ref{fig:lwall}). If a bar narrower than the FWHM of the PSF is smeared out to the PSF size, this could introduce a bias such that strength measurements including bar widths may be affected by these broadened bars. In the GZBL-COSMOS data examined here, this effect is mild, but still present. Thus, in the analysis below, we further restrict the \barmass\ sample to those bars where the width measurement is $\mwbar > 1.5$~kpc, which is resolved by the $HST$ ACS PSF at all redshifts. 
This reduces the sample size to \mcgzhcosmosdiskbarres\ barred galaxies. We have verified that our qualitative conclusions below do not change for reasonable choices of resolution threshold. Given that the bar length exceeds the width by definition, enforcing resolution via bar widths is more conservative than requiring well resolved lengths.

\subsection{Overall trends with galaxy properties}\label{sec:barpropgalprop_overallsubsec}

Figure \ref{fig:overallbarprops} presents the trends of four bar properties: absolute bar length (\lbar , in kpc), relative bar length (\rrel), relative bar width (\wrel), and length-to-width ratio (\lwrat), as functions of redshift ($z$), stellar mass (\mstar ), and offset from the star-forming sequence midpoint (\dlogsfr ). The length-to-width ratio also serves as a measure for bar strength that correlates well to other measures \citep{Laurikainen2002a,Laurikainen2002b,laurikainen07}, where larger ratios equate to stronger bars. 
Dashed lines on the figure trace the width of the distribution, enclosing 68\% of barred galaxies in that bin. We also compute binned median values and show these on the figure. Variances on the median are generally smaller than the line width, so we estimate uncertainties on medians by varying bin edges within a reasonable tolerance (drawn randomly from a uniform distribution 20\% of the bin width in each quantity), resampling the bins 1,000 times. The shaded regions around the binned medians represent the 95\% confidence interval on the median determined via this method.\footnote{These confidence intervals/uncertainties are usually larger than the formal variance on the median, but are still relatively small (sometimes $<1$\%). In the rest of the discussion, we sometimes quote median values with more decimal places than we would ordinarily consider significant, in order to quote non-zero uncertainties on the median.}

The median physical bar length of the overall sample is $\langle \mlbar \rangle_{\rm med} = \barmassLmedres$~kpc. The top left panel of Figure \ref{fig:overallbarprops} also shows a dotted line representing the physical size of twice the FWHM of the \emph{HST} ACS PSF, an estimate of the minimum length we can resolve in the \barmass\ sample. We do not find significant evolution of the median physical bar length (represented by the solid black line) within this dataset. This observation agrees with \cite{kim21}, who use a sample of 379 galaxies within $0.2 < z < 0.84$ to show that there is similarly no significant evolution of physical bar length within their redshift range. Our median \lbar\ value agrees with the mean value of \citeauthor{kim21} within their standard deviation; both are about twice as long as the mean physical length of bars measured with \emph{JWST} at $1 \leq z \leq 4$ \citep{LeConte2025}, and significantly larger than the minimum resolvable bar length at all redshifts.

Within our sample, the median relative length \rrel\ also shows no significant evolution with redshift across all redshift bins ($\langle \mrrel \rangle_{\rm med} = \rrelmedzOneres_{-\rrelmedzOneDlores}^{+\rrelmedzOneDhires}$ in the lowest-redshift bin versus $\langle \mrrel \rangle_{\rm med} = \rrelmedzFourres_{-\rrelmedzFourDlores}^{+\rrelmedzFourDhires}$ in the highest-redshift bin). This is consistent with the mean and standard deviations of the similar-redshift sample of \citeauthor{kim21} and the higher-redshift sample of \citeauthor{LeConte2025}, implying a stable ``typical'' relative bar length across at least 9 Gyr of cosmic time, over which period the size of disk galaxies evolves significantly (which reconciles a factor of two evolution in physical size but no evolution in relative size). 

The median bar width \wrel\  also appears marginally stable, with the relative width in the lowest redshift bin ($\wrelmedzOneres_{-\wrelmedzOneDlores}^{+\wrelmedzOneDhires}$) just consistent with that of the highest redshift bin ($\wrelmedzFourres_{-\wrelmedzFourDlores}^{+\wrelmedzFourDhires}$) within the 95\% uncertainties. While neither the slight decrease in median \rrel\ nor the slight increase in median \wrel\ with $z$ is statistically significant, the length-to-width ratio (\lwrat) shows a decrease with increasing redshift on the median of approximately 13\% ($\lwratmedzOneres_{-\lwratmedzOneDlores}^{+\lwratmedzOneDhires}$ to $\lwratmedzFourres_{-\lwratmedzFourDlores}^{+\lwratmedzFourDhires}$), within the 95\% uncertainties. This may imply that bars typically develop as slightly weaker instabilities at higher redshifts, and strengthen over time. This is consistent with bars in the population becoming stronger over time as long-lived bars absorb angular momentum \citep{athanassoula13}. It is also potentially consistent with the slightly lower ellipticities of bars at $z > 1$ measured by \citet{LeConte2025}, where their average corrected ellipticity of $\approx 0.5$ would correspond to a length-to-width ratio of $\approx 2$. However, the inherent uncertainties that \citeauthor{LeConte2025} carefully describe in their measurements, as well as differences in methods and mass completeness between studies, mean that further work would be required to quantify the evolution of bar strength across these epochs. There is also considerable range in values of both relative length and width among populations at all redshifts, including those sampled in this work. 

In a recent analysis of the Auriga simulations \citep{Grand2017,Grand2019}, \citet{Fragkoudi2025} find that most bars initially develop at shorter relative lengths, then lengthen as they age; however, bars that form at higher redshifts (including the high end of our redshift range) tend to form at longer relative lengths, which then stay at approximately the same \rrel\ as the system ages. The relative consistency in lengths across the redshift range of our sample may be consistent with this combination of bars strengthening as they age, but a higher proportion of newer, weaker bars at higher redshifts. \citeauthor{Fragkoudi2025} define the bar strength of their simulated galaxies as the $m=2$ Fourier mode of the surface density, $A_2$, and find that bars that develop early remain at similar strengths throughout their evolution, whereas bars that develop later may slowly increase their strength (their Figure 3). We cannot determine the bar ages of this sample \citep[although perhaps further data could enable this, \eg using the method described by][]{deSa-Freitas2025}, so we cannot separate the general trend in the lower left panel of Figure \ref{fig:overallbarprops} into younger and older bars. However, the trend is shallow enough and the differences in method and galaxy mass coverage are complex enough that our results may be consistent with those of \citeauthor{Fragkoudi2025} 

Note that this trend in median \lwrat\ values (and all trends reported here), while significant compared to the bootstrapped uncertainty on the median, is small compared to the overall width of the distribution of \lwrat\ values: the changes are subtle compared to the variety within the population. \cite{kim21} found that mean bar strength has no significant evolution with redshift, over approximately the same redshift range we probe. We would note that there is a slight decrease in their mean bar strength with increasing redshift in Figure 7 of \citeauthor{kim21}, but as variances on means are generally much larger than variances on medians, their claim of a lack of significance in the mean trend is consistent with our finding of a change in the median value that is real, but small compared to the distribution width. Differences between the bar strengths of \cite{kim21} and the bar strength results of this study may also be due in part to the difference in our bar strength definitions, since \cite{kim21} defines the bar strength using two methods: Fourier amplitude ($A_2$) and an estimate of torque parameter ($Q_b$), rather than our method of simply taking the ratio \lwrat. 


We find more notable trends when examining these 4 properties (\lbar, \rrel, \wrel, \lwrat) as functions of stellar mass compared to the  slight (or no) trends with redshift. The median physical bar length increases by 37\% across the full range of stellar mass (from $\LmedMOneres_{-\LmedMOneDlores}^{+\LmedMOneDhires}$ to $ \LmedMFourres_{-\LmedMFourDlores}^{+\LmedMFourDhires}$ kpc from lower to higher mass). The direction of this trend is expected given that higher-mass galaxies tend to have larger bars \citep{hoyle11,kim21}. While there is a slight increase in the slope of the correlation at higher stellar masses, we do not observe the same strongly broken trend as \citet{erwin19}, who found that bar lengths in lower-redshift galaxies have a nearly constant mean length in galaxies with mass up to just above $10^{10} ~\mmsun$, with lengths increasing steeply with mass in more massive galaxies. Interestingly, we observe a median trend between physical bar length and galaxy stellar mass more consistent with that of \citeauthor{erwin19} if we include bars with narrow widths that may not be well resolved at these redshifts (\ie if we skip the conservative resolution cuts described at the start of this Section). This suggests that the change in the shape of the trend may be largely due to narrower bars, to which \citet{erwin19} was far more sensitive owing to the high physical resolution of the dataset used in that study.

The stellar mass at which we observe a slight additional increase in bar length is slightly higher than the break mass found by \citeauthor{erwin19}; both are similar to the mass at which other bar properties transition \citep[\eg][]{nair10b,kruk17} given the evolution of the galaxy mass function between those studies' and our samples' redshifts \citep{Weigel2016,Driver2022,Shuntov2025}. 
This mass also corresponds to the mass range beyond which quiescence tends to onset and the ``red sequence'' begins to dominate the galaxy population versus the ``blue cloud'' \citep{Kauffmann2003,smethurst15}.

\begin{figure*}
\begin{center}
\includegraphics[width=0.85\textwidth]{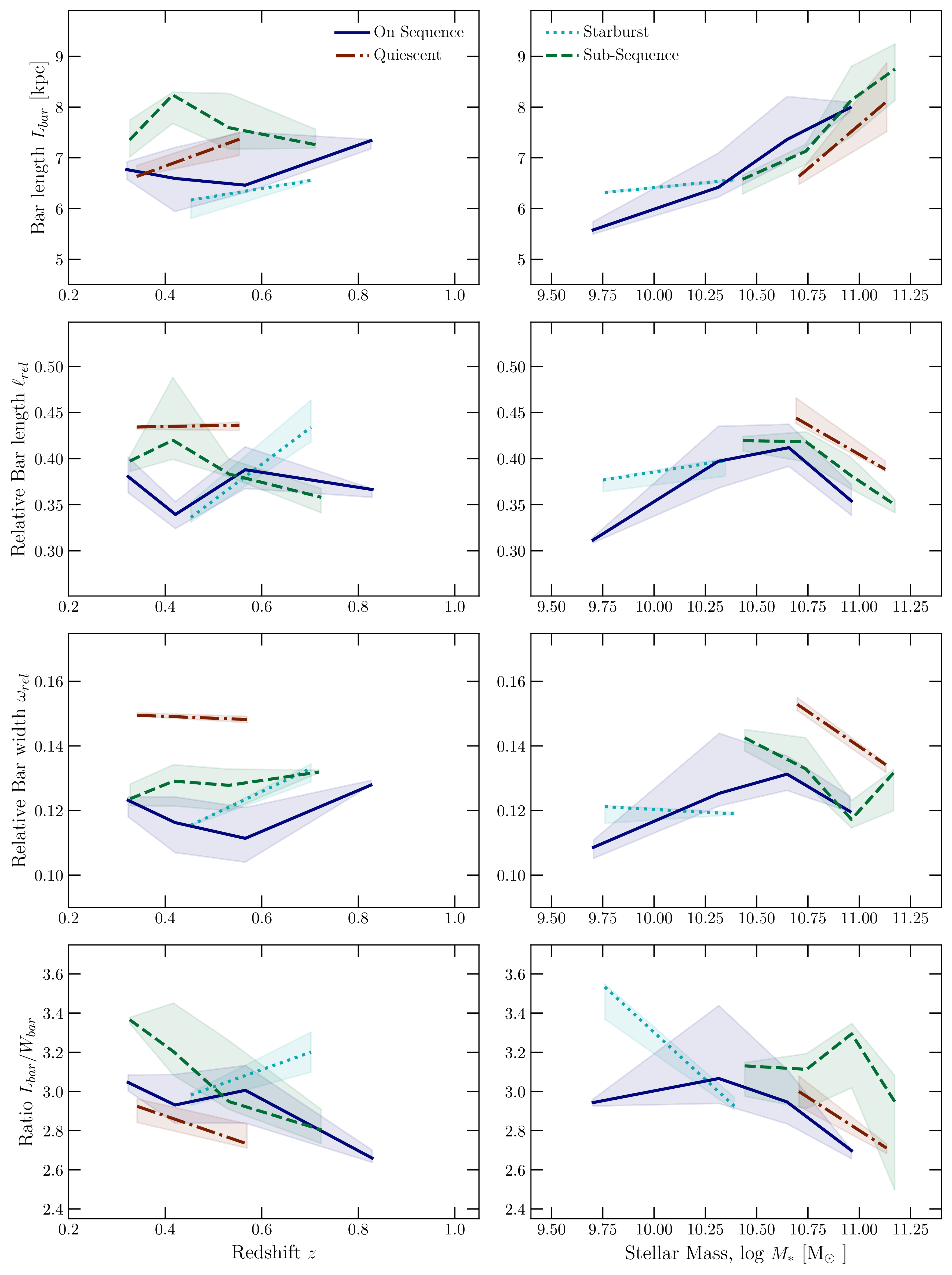}
\caption{Bar length \lbar\ (top), relative bar length \rrel\  (upper middle), relative bar width \wrel\ (lower middle), and bar strength (bottom) versus redshift $z$ (left) and stellar mass \mstar\ (right) within the subset of the \barmass\ sample with $\mwbar > 1.5$~kpc. The sample has been split by star-forming status, with each galaxy assigned to be starburst (cyan dotted), on the star-forming sequence (blue solid), sub-sequence (green dashed), or quiescent (red dot-dashed) based on its position on the SFR-\mstar\ diagram relative to the centre of the star-forming sequence at its stellar mass and redshift (\dlogsfr ). Many \dlogsfr -based differences with redshift disappear when comparing galaxies at the same stellar masses across the star-forming categories, but there are still some statistically significant median differences between galaxies in different star-formation categories at a given stellar mass and a given redshift.
}
\label{fig:barprops_galaxyprops_sfr}
\end{center}
\end{figure*}

The median relative bar length increases by about 30\% up to $\lmstartxt \sim 10.25$, then decreases again towards higher stellar masses. The difference between this trend and the trend in physical bar length illustrates the importance of considering the overall galaxy size in consideration of bar lengths: physical bar length increases faster than galaxy disk size with mass, except at the highest masses in the sample, where the 90-per-cent radii of $\mmstar \sim 10^{11}~\mmsun$ disks are large enough that the relative bar size is smaller even though the bar is also physically longer. Relative bar width shows a similar median trend, increasing by about 17\% to the same mass and falling back slightly at the highest masses. The median ratio \lwrat\ with stellar mass is consistent with no change (\eg $\lwratmedMOneres_{-\lwratmedMOneDlores}^{+\lwratmedMOneDhires}$ in the lowest-mass bin versus $\lwratmedMFourres_{-\lwratmedMFourDlores}^{+\lwratmedMFourDhires}$ in the highest), indicating no significant change in bar strength with increasing stellar mass. 

\citet{jang25}
predicts that bars in lower-mass disk-dominated galaxies should be weaker than bars in higher mass galaxies, directly contradicting our finding here. However, \citeauthor{jang25} focused on bar development in the absence of bulges, a restriction we have not imposed on our data. Most of our disk galaxies have at least some bulge. Therefore, this discrepancy may reflect a dependence on bulges in the formation and evolution of bars, as well as the uncertainty around observational proxies derived from images compared to the more direct dynamical measurements that are possible when using simulated data.

Our findings generally align well with those of past observational studies. At low redshift, bars tend to be longer (both in physical size and relative to the disk size) in more massive \citep{erwin05b}, redder \citep{hoyle11} galaxies. \citet{kim21} similarly find that bar lengths scale with host galaxy mass at the same redshifts as this work, giving further support to the fact that bars and disks co-evolve with each other and maintain their relative proportions. In studies of very nearby galaxies, bar lengths tend to vary with morphological T-type in the same way as luminosity varies with morphology \citep[\eg][]{erwin05b,menendezdelmestre07}, so our higher-redshift results are qualitatively similar to lower-redshift results. A more detailed comparison with these studies is complicated by the very different selection methods used in studies of local galaxies, which we cannot replicate (\eg sensitivity to much smaller bars using space-based instruments to observe the local Universe, as in both \citealt{erwin05b} and \citealt{menendezdelmestre07}). 
The lack of change in our indicator of bar strength with stellar mass is qualitatively consistent with low-redshift studies which use similar strength metrics \citep{Whyte2002,laurikainen07}.


The trends with \dlogsfr\ highlight the possible presence of structural differences in bars as a function of star-forming activity. We find that \lbar\ decreases as \dlogsfr\ increases (from $\LmedSFROneres_{-\LmedSFROneDlores}^{+\LmedSFROneDhires}$~kpc in the combined Quiescent category to $\LmedSFRFourres_{-\LmedSFRFourDlores}^{+\LmedSFRFourDhires}$~kpc in the Starburst category), and the same is true for both \rrel\ (from $\rrelmedSFROneres_{-\rrelmedSFROneDlores}^{+\rrelmedSFROneDhires}$ to $\rrelmedSFRFourres_{-\rrelmedSFRFourDlores}^{+\rrelmedSFRFourDhires}$) and \wrel\ (from $\wrelmedSFROneres_{-\wrelmedSFROneDlores}^{+\wrelmedSFROneDhires}$ to $\wrelmedSFRFourres_{-\wrelmedSFRFourDlores}^{+\wrelmedSFRFourDhires}$). Galaxies lying on or above the star-forming sequence host bars that are shorter and narrower than those in quenching or quiescent galaxies. 

The decrease in relative width is slightly larger than the decrease in relative length across the star-forming categories. These trends combine such that the median \lwrat\ \emph{increases} very slightly from lowest to highest \dlogsfr, by about 7\% (from $\lwratmedSFROneres_{-\lwratmedSFROneDlores}^{+\lwratmedSFROneDhires}$ in the combined Quiescent category to $\lwratmedSFRFourres_{-\lwratmedSFRFourDlores}^{+\lwratmedSFRFourDhires}$ in the Starburst category). In our definition, larger ratios correspond to stronger, more elongated bars. This means that galaxies above the star-forming sequence tend to host physically shorter but slightly stronger bars, while galaxies in the combined ``Quiescent'' category host weaker, rounder bars. The slope of the change is not continuous across star-forming categories, and differences between any two categories are small. As with all other trends we observe, the statistically significant changes in median values are much smaller than the scatter within the overall population. In the lower right panel of Figure \ref{fig:overallbarprops}, the upper dashed line indicating the 84th percentile of the distributions also trends toward stronger bars with higher \dlogsfr , but there is no clear trend in the lower line tracing the 16th percentile of the distribution.

Our results connecting bar structure to \dlogsfr\ add to the broader context in which bars are tied in complex ways to the regulation of star formation.
For example, \citet{geron24} combined a bar strength definition from Galaxy Zoo DESI classifications (in which ``strong" and ``weak" bars are distinguished by their visual prominence) with spatially resolved spectroscopy to show that strong bars in star-forming galaxies enhance star formation in their centers \citep[an effect predicted by theoretical work, \eg][]{sellwood14} and just beyond the bar-ends, while suppressing it along the bar arms. They further found that this effect is strongest for slow, strong bars, suggesting that bar kinematics as well as morphology influence how and where star formation occurs. \citet{Mengistu2026} further found that bars' effect on host galaxy star formation is strongest on shorter timescales versus other secular processes. Both analyses were limited to nearby galaxies ($z < 0.06$) owing to the observational constraints involved. 

Our photometric finding that galaxies above the star-forming sequence host shorter but proportionally slightly stronger bars is complementary to the results from \citet{geron24}, potentially illustrating how \citeauthor{geron24}'s more localized star formation patterns may show up in the more global view of star formation patterns across the full range of \dlogsfr. However, given that \dlogsfr\ is also correlated with stellar mass (\eg Figure \ref{fig:sfrmass_ratio}), and that both star formation rate and stellar mass evolve with redshift, it may be useful to disentangle these quantities further while also examining bar properties.

\subsection{A closer look at bar properties with star formation}\label{sec:barpropgalprop_SFsubsec}

In Figure \ref{fig:barprops_galaxyprops_sfr}, we investigate how bar structure varies within \dlogsfr\ categories by examining median \lbar, \rrel, \wrel, and \lwrat\ ratio as functions of redshift and stellar mass (as in Figure \ref{fig:overallbarprops}), but now separated by star formation status. The 95\% confidence intervals on median values use the same bootstrapping method as the uncertainties in Figure \ref{fig:overallbarprops}. 
As expected, there are not enough bars at low mass or high redshift hosted in quiescent galaxies to confidently discuss median trends; likewise for starburst galaxies at both redshift extremes and at high stellar mass. 

At the lower redshifts in the sample ($z \lesssim 0.6$), galaxies with lower star formation activity (``Sub-Sequence'' and ``Quiescent'') tend to host bars that are both longer and wider than galaxies on or above the star-forming sequence. In terms of physical length, the Sub-Sequence galaxies (which are in the region of the $SFR-\mmstar$ plot known as the ``green valley'') are the only category that shows a significant difference in median \lbar\ versus galaxies on the star-forming sequence. Bars in quenching/green-valley galaxies are $\approx 0.6_{-0.4}^{+0.4}$~kpc longer than bars in galaxies on the star-forming sequence at $z \approx 0.32$, and $\approx 1.6_{-0.7}^{+0.8}$~kpc longer than those in galaxies on the sequence at $z \approx 0.42$. 

This redshift difference represents an evolutionary time of 0.84 Gyr, and the time between the median redshift in the $z \approx 0.42$ bin and the next-highest redshift bin (with a combined median across both categories of $z \approx 0.55$) is 0.94 Gyr. Quenching on fast ($<1$ Gyr) and intermediate ($<2$ Gyr) timescales does occur in disk galaxies \citep[\eg][]{smethurst15}, and bars can affect their galaxies' overall star formation on these timescales \citep[\eg][]{Mengistu2026}. Bar strength can also vary significantly in individual galaxies over timescales less than 1 Gyr \citep{Fragkoudi2025}, so it is possible for galaxies to both transition from star forming to quenching and substantially change their bar properties across redshift bins sampled here. The relative bar lengths between ``On Sequence'' and ``Sub-Sequence'' are also different at lower redshifts, more so than the relative widths, such that the ratio \lwrat\ is significantly higher for sub-sequence galaxies at the lowest redshifts in the sample than for any other category ($\lwratmedSubSeqzOneres_{-\lwratmedSubSeqzOneDlores}^{+\lwratmedSubSeqzOneDhires}$ for sub-sequence galaxies versus $\lwratmedOnSeqzOneres_{-\lwratmedOnSeqzOneDlores}^{+\lwratmedOnSeqzOneDhires}$ for on-sequence galaxies at $z \approx 0.32$). 

Comparing the ``Sub-Sequence'' and ``On Sequence'' bar lengths as a function of stellar mass adds further context: the masses of barred disk galaxies that are quenching are typically higher than those of barred star-forming disks, and the trend of increasing physical bar lengths with increasing stellar mass is consistent between both sub-sample. The trend in \lwrat\ with stellar mass is partially consistent where the quenching and on-sequence galaxies overlap. A galaxy on the star-forming sequence that begins quenching at a stellar mass of $\mmstar = 10^{10}\ \mmsun$ could potentially continue to grow while its SFR falls slightly below the sequence and reach the typical mass of the lowest-mass bin for sub-sequence galaxies in Figure \ref{fig:barprops_galaxyprops_sfr}, but it would need most of the time elapsed between the highest and lowest redshift bins in our sample to do so. Together, these indicate that the differences between these subsamples as a function of redshift are likely at least partially due to the fact that quenching tends to onset in higher-mass barred galaxies. 

However, mass effects alone likely do not explain this. Barred galaxies in the combined Quiescent category have typical stellar masses at least as high as the Sub-Sequence barred galaxies, and their bar lengths are lower than both on-sequence and sub-sequence galaxies at the same masses. The relative lengths of quiescent bars are higher than those of bars in galaxies on the star-forming sequence, though not necessarily than those of bars in galaxies that are still quenching. The relative bar widths in quiescent galaxies are significantly higher than bars in any of the other star-forming categories. This results in \lwrat\ ratios substantially below those of bars in sub-sequence galaxies at $z \lesssim 0.6$, and consistently slightly below (though only marginally different from) those of bars on the star-forming sequence. At the higher stellar masses where we sample Quiescent, Sub-Sequence, and On Sequence barred galaxies, there is generally a progression whereby the median relative lengths and widths of bars in star-forming galaxies are smaller than those of quenching galaxies at a given stellar mass, which are themselves smaller than those of quenched galaxies at a given stellar mass. These differences mostly divide out when examining \lwrat\ as a function of stellar mass: \lwrat\ between star formation categories within the same overlapping mass range has relatively high uncertainties and does not show an obvious progression, nor does it preserve the ordering of categories.

At higher redshifts ($z \gtrsim 0.6$), most differences between \dlogsfr\ categories are not statistically significant (with the caveat that we do not have enough quiescent barred galaxies at higher redshifts to reliably determine median bar properties). The only exception is the Starburst category, whose bars are slightly longer relative to their (lower mass and more compact) host galaxies than those in on-sequence or sub-sequence galaxies. The right column of Figure \ref{fig:barprops_galaxyprops_sfr} shows that this is driven by lower-mass galaxies. Bars in starburst galaxies are longer, wider, and stronger (via \lwrat ) than galaxies on the star-forming sequence, but only at the lowest masses we sample (the median mass of a starburst galaxy in this bin is $\lmstartxt = 9.76$).

The offsets in bar properties between galaxies of the same mass and redshift but different star-forming status indicate that bar structure may be directly connected to a galaxy's level of star formation and consequently, its evolutionary state. However, our study echoes the results of many others \citep[\eg][]{Fraser-McKelvie2020a,Fraser-McKelvie2020b,geron24,Mengistu2026} in finding that the link between bar evolution and the regulation of star formation is likely to be complex.

For example, it is well known that galaxy interactions--such as mergers and close fly-bys--can trigger bursts of star formation \citep[\eg][]{Barton2000,DiMatteo2007,Shah2022,ORyan2025a}. These episodes tend to occur over relatively short dynamical timescales and produce characteristic and asymmetric morphological features \citep[\eg][]{lotz04a,lotz08b,lotz08a,Ellison2008,Patton2013}. Interaction-triggered bars are more likely to form both longer and stronger bars \citep{Fragkoudi2025}, although whether they are short- or long-lived depends somewhat on the dynamical stability of the pre-interaction disk \citep{kraljic12}. 

However, merging systems are morphologically excluded from this study, so those events cannot explain the trends we observe unless the merger-induced differences persist until after visual merger signatures have relaxed \citep{lotz08b}. This is certainly possible for some galaxies in the sample, but it is unlikely to explain all the effects we see. To understand the internal set of processes likely dominating our results, \citet{geron24} provide a useful point of comparison. Their results outline a sequence of events echoing the ``bar continuum'' proposed by \citet{Geron2021}. Combining that proposal with our results and those of the simulations mentioned above is consistent with a scenario in which a strong bar forms, drives gas inward along its length, and boosts star formation in the center and near the bar-ends. Over time, this contributes to global quenching of the galaxy, leading to somewhat weaker, more extended bars in quiescent systems. This timeline implies multiple timescales for bar-driven evolution, most of which differ from the shorter and more transient bursts of star formation expected from fly-by interactions. 

Enhanced AGN activity would also be expected to be associated with the presence of a relatively young bar in the above scenario. While this was an unsettled debate for many years \citep[\eg][]{martini03,lee12,galloway15,cheung15}, there is now clear evidence for an increase in the AGN fraction in barred galaxies \citep{Garland2024,LaMarca2026}. The enhancement is stronger in more strongly barred systems \citep{Garland2024}, and is more associated with the lower- and moderate-luminosity AGN that are more likely to be triggered via secular processes than the higher-luminosity growth seen as a result of interactions and mergers \citep{LaMarca2026}. Examining the joint role of AGN and star-formation status in bar evolution is out of scope for this work, but this will be possible in future, larger samples that retain their statistical power for examining AGN fractions in different star-formation subcategories of AGN host galaxy.

The redshift range of the \barmass\ sample spans approximately 5 Gyr of cosmic time. Our findings of modest differences in median bar properties as functions of redshift and mass, with some complexities in the differences between barred galaxies at different levels of star-formation, provide observational evidence consistent with the picture in which bar growth, gas inflows, and quenching proceed together via both secular processes at multiple timescales and more rapid externally-driven events.

%
%
\section{Summary}\label{sec:summary}
%
%

In this study, we investigate how bar presence and structure relates to star formation activity in disk galaxies across cosmic time. Using bar length and width measurements and galaxy classifications from Galaxy Zoo, we analyze a mass-complete sample of barred and unbarred galaxies in the COSMOS survey over the redshift range $0.25 < z < 1$ and with stellar masses $\lmstartxt \geq 9.5$. Our analysis focuses on whether barred galaxies differ systematically from their unbarred counterparts in terms of star-forming properties, and how bar lengths and other structural properties evolve with stellar mass, redshift and distance from the star-forming sequence. The key findings of this work are as follows:

\begin{enumerate}

    %

    \item After controlling for differences in redshift and stellar mass distributions, the star-formation rates of barred galaxies are significantly lower than those of unbarred galaxies, on average. In particular, at stellar masses $\lmstartxt \gtrsim 10$, barred galaxies are more likely to occupy the ``green valley'', just below the star-forming sequence, or be quiescent. This is consistent with bars being associated with relatively slow quenching processes, and with the presence of a bar contributing to the observed scatter in the star-forming sequence.

    \item The median bar length relative to disk length does not evolve over the redshifts we probe, but the median length-to-width ratio, a proxy for bar strength, shows a trend with redshift, decreasing 13\% from the lowest to highest redshift of the sample. The median bar at higher redshift is slightly weaker/less elongated than at lower redshift. This is broadly consistent with recent theoretical work. 

    \item Median physical bar length (in kpc) increases with stellar mass for galaxies, by nearly 40\% across our stellar mass range. This finding is similar to that of previous studies of lower-redshift barred galaxies.
    
    \item In relative terms, bars are at their proportionally longest \emph{and} widest at  $\lmstartxt \sim 10.25$, and both quantities change with mass such that the median bar strength is consistent with no significant change across the mass range of our sample. 

    \item Median bar lengths and widths also change as a function of star-formation rate, such that quiescent and quenching galaxies have the longest and widest bars (relative to their host disks) compared to galaxies both on and above the star-forming sequence.  Given that quiescent and quenching (sub-sequence) galaxies also predominantly have higher stellar masses, this finding is generally consistent with bars contributing to the secular decline of star formation in massive disk galaxies over time. However, more rapid quenching due to processes such as galaxy fly-bys (which can cause bar formation) is not ruled out, and some of the complexities in the differences in median bar properties with redshift and stellar mass indicate that more rapid quenching timescales are likely also present in our sample.

    \item Starburst galaxies, which mainly populate the low-mass end of the overall population, show an increase in physical and relative length, as well as length-to-width ratio, compared to galaxies at the lowest masses in the sample that are on the star-forming sequence.
    
    \item While there are complexities in bar properties with redshift and stellar mass after separating the barred population into star-forming categories, the persistence of an overall sequence in relative bar lengths and widths from actively star forming to quiescent systems suggest that bar properties evolve alongside and participate in overall galaxy evolution.

    %
  
\end{enumerate}

Together, these results support a picture in which bar-driven secular processes play a measurable role in the quenching of disk galaxies. Past theoretical work has shown that bars offer a structural route to quenching, and past observations of bar and galaxy kinematics have found clear indications of bar-associated changes to local star formation. In this context, our results imply that bars at least partially drive and record a galaxy's transition from actively star-forming to quiescent. 

While all the median differences and trends we report here are statistically significant, it is important to note that the overall variations of the population (or sub-population) distributions are generally much wider than the median trends. There is still potentially much to learn about disk galaxy evolution within this scatter.

In recent years, the study of bar features in disk galaxies has moved forward via a combination of advanced simulations and multiple observational techniques, including imaging studies of larger samples and targeted dynamical work. Looking forward, a new generation of high-resolution surveys (\eg \emph{JWST}-driven surveys such as CEERS, \citealt{Finkelstein2025}, and COSMOS-Web, \citealt{Casey2023}; as well as surveys with \emph{Euclid}, \citealt{EuclidCollaboration2025}, and \emph{Roman}, \citealt{Spergel2015,Akeson2019}), as well as Adaptive-Optics-assisted ground-based surveys using instruments such as MUSE, promise to bring us spatially-resolved star-formation insights on large samples of barred and unbarred disk galaxies at rest-frame optical and infrared wavelengths across many Gyr of cosmic time. Among other discoveries, these will determine whether the global trends we observe here align with the smaller-scale bar effects described by current kinematic studies.

%
%
\section*{Data Availability}
%
%

The full catalogue from the Galaxy Zoo Bar Lengths project, including bar measurements across all \emph{HST} surveys covered and links to all subject images classified, will be available at \url{https://data.galaxyzoo.org} and \url{https://doi.org/10.5281/zenodo.19614348} upon publication of this manuscript.

%
%
\begin{acknowledgments}
%
%
%

This work has taken place over many years and through many author career stages. THS appreciates and acknowledges the support from Spelman College, the Center for Astrophysics and Space Sciences (now the Department of Astronomy and Astrophysics) at the University of California San Diego, the Morehouse-Spelman Bridge Program, the STARS Program, the University of California-Historically Black College and University (UC-HBCU) Fellowship, and the support of Brooke Simmons and Adam Burgasser. 

BDS acknowledges past support from Balliol College, Oxford through the Henry Skynner Junior Research Fellowship, and from the National Aeronautics and Space Administration (NASA) through Einstein Postdoctoral Fellowship Award Number PF5-160143 issued by the Chandra X-ray Observatory Center, which is operated by the Smithsonian Astrophysical Observatory for and on behalf of NASA under contract NAS8-03060. BDS acknowledges current support from a UK Research and Innovation Future Leaders Fellowship [grant number MR/T044136/1] and its renewal [grant number MR/Z000076/1].

ILG has received support from the Czech Science Foundation Junior Star grant no. GM24-10599M. RJS gratefully acknowledges support through the Royal Astronomical Society Research Fellowship. 

This publication uses data generated via the \url{Zooniverse.org} platform, development of which is funded by generous support, including a Global Impact Award from Google, and by a grant from the Alfred P. Sloan Foundation. We are particularly grateful the GZBL volunteers whose contributions enabled the bar measurements used in this work. Each registered volunteer contributor is acknowledged as a team member at \url{https://www.zooniverse.org/projects/vrooje/galaxy-zoo-bar-lengths/about/team}. The Zooniverse-Panoptes {\tt project\_id} of the Galaxy Zoo Bar Lengths citizen science project is 3. 

This research is in part based on observations made with the NASA/ESA \emph{Hubble Space Telescope}, obtained at the Space Telescope Science Institute, which is operated by the Association of Universities for Research in Astronomy, Inc., under NASA contract NAS5-26555. The public data used in Galaxy Zoo Bar Lengths was collected as part of observing programmes 
GO-9425, 
GO-9500, 
GO-9583, 
GO-9822, 
GO-10134, 
GO-12060, 
and GO-12099. 
We are grateful to the GOODS, GEMS, COSMOS, AEGIS, and CANDELS project teams for their efforts in providing high-level science products to the public.

The Dunlap Institute is funded through an endowment established by the David Dunlap family and the University of Toronto.

\end{acknowledgments}

\begin{contribution}

THS led the data analysis, literature review, and wrote the majority of the manuscript.
BDS created and was PI of the \emph{Galaxy Zoo Bar Lengths} citizen science project, aggregated the volunteer classifications, wrote the sections of the manuscript describing the catalogue, and contributed to the writing and editing of some other parts of the manuscript.
KLM is the PI of the Galaxy Zoo project, and CL was the PI of the Zooniverse (and Galaxy Zoo) at the time of the launch of the GZBL project. They and all other authors contributed ideas and discussion to the project. This includes Zooniverse volunteers EB, SB, MLP, and JW, all of whom collaborated with the other authors to make key decisions about the design and primary goals of the GZBL project.

\end{contribution}

%
\facilities{HST(ACS), HST(WFC3)}

\software{
          In addition to the software cited in the main text, this research made extensive use of the the Python packages {\tt astroPy} \citep{astropy13,AstropyCollaboration2018,AstropyCollaboration2022}, {\tt matplotlib} \citep{Hunter2007}, {\tt numpy} \citep{Harris2020}, {\tt pandas} \citep{mckinney-proc-scipyconf-2010-pandas} v1.4.2 \citep{jeff_reback_2022_6408044}, the Python Image Library \citep{clark2015pillow}, and {\tt scipy} \citep{Virtanen2020}. We also used NASA's ADS/SciX services and Cornell's ArXiv. This publication also used a widget form of the JavaScript Cosmology Calculator \citep{wright06,rsimpson13} and the Tool for Operations on Catalogues And Tables (TOPCAT; ~\citealt{Taylor05,Taylor2011})\footnote{\url{http://www.star.bris.ac.uk/~mbt/topcat/}}.
          }


\bibliography{refs,gzbl_refs,softwarecitations}{}
\bibliographystyle{aasjournalv7}



\end{document}